\documentclass[10pt,twocolumn,letterpaper]{article}

\usepackage[pagenumbers]{cvpr} % To force page numbers, e.g. for an arXiv version

\definecolor{cvprblue}{rgb}{0.21,0.49,0.74}
\usepackage[pagebackref,breaklinks,colorlinks,allcolors=cvprblue]{hyperref}

\usepackage{multirow}
\usepackage{multicol}
\usepackage[ruled,lined,linesnumbered]{algorithm2e}
\usepackage[dvipsnames]{xcolor}

\graphicspath{{figs/}}
\newcommand{\green}[1]{{\textcolor{ForestGreen}{\textbf{#1}}}}

\newcommand{\gold}{\raisebox{-0.2ex}{\includegraphics[height=1.0\ht\strutbox]{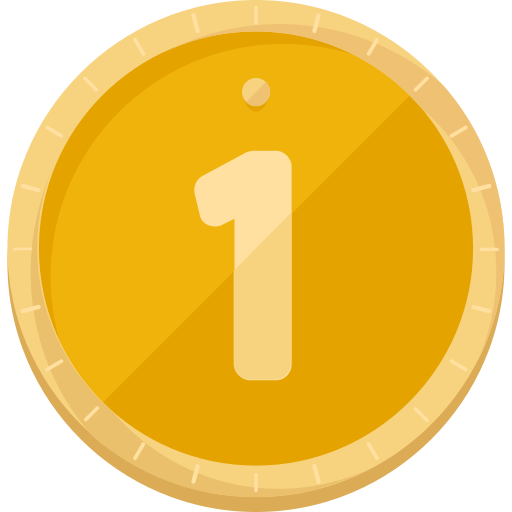}} }
\newcommand{\silver}{\raisebox{-0.2ex}{\includegraphics[height=1.0\ht\strutbox]{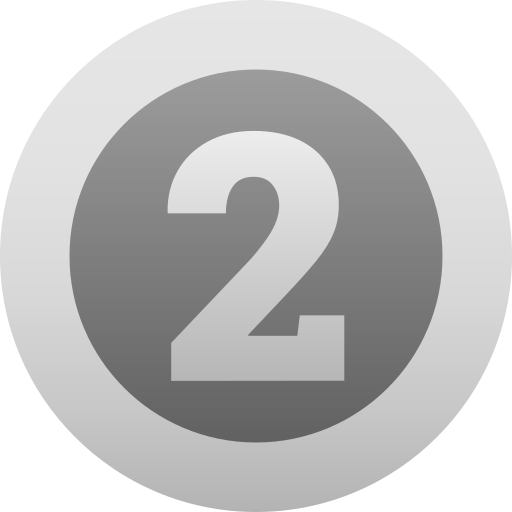}} }

\ifx\figurename\undefined \def\figurename{Figure}\fi
\renewcommand{\figurename}{Fig.}
\renewcommand{\paragraph}[1]{\textbf{#1} }

\newcommand{\Sect}[1]{Sec.~\ref{#1}}
\newcommand{\Fig}[1]{Fig.~\ref{#1}}
\newcommand{\Tbl}[1]{Tbl.~\ref{#1}}
\newcommand{\Eqn}[1]{Eqn.~\ref{#1}}

\newcommand{\Alg}[1]{Algo.~\ref{#1}}

\newcommand{\specialcell}[2][c]{\begin{tabular}[#1]{@{}c@{}}#2\end{tabular}}

\def\cI{{\textsc{I}}}
\newcommand{\cL}{\mathcal{L}}
\def\cG{{\textsc{G}}}
\def\cP{{\textsc{P}}}
\def\cHP{{\textsc{HP}}}
\def\cCR{{\textsc{CR}}}

\newcommand{\proj}{\textsc{Seele}\xspace}
\newcommand{\mode}[1]{\underline{#1}\xspace}

\def\paperID{5338} % *** Enter the Paper ID here
\def\confName{CVPR}
\def\confYear{2026}

\title{\proj: A Unified Acceleration Framework for Real-Time Gaussian Splatting on Mobile Devices}

\author{
Xiaotong Huang$^1$\footnote{Equal contribution.} \quad He Zhu$^1$\footnotemark[1] \quad Zihan Liu$^1$$^2$ \quad Weikai Lin$^3$   \quad Xiaohong Liu$^1$ \quad Zhezhi He$^1$ \\
Jingwen Leng$^1$$^2$ \quad Minyi Guo$^1$$^2$ \quad Yu Feng$^1$\footnote{Corresponding Author.} \\
$^1$Shanghai Jiao Tong University \quad $^2$Shanghai Qi Zhi Institute \quad $^3$University of Rochester\\
\vspace{0.5em}
\small Project Page: \url{http://seele-project.netlify.app}
}

\begin{document}
\maketitle
\begin{abstract}

3D Gaussian Splatting (3DGS) has become a crucial rendering technique for many real-time applications. 
However, the limited hardware resources on today's mobile platforms hinder these applications, as they struggle to achieve real-time performance. 
In this paper, we propose \proj, a general framework designed to accelerate the 3DGS pipeline for resource-constrained mobile devices.

Specifically, we propose two GPU-oriented techniques: hybrid preprocessing and contribution-aware rasterization.
Hybrid preprocessing alleviates the GPU compute and memory pressure by reducing the number of irrelevant Gaussians during rendering.
The key is to combine our view-dependent scene representation with online filtering. 
Meanwhile, contribution-aware rasterization improves the GPU utilization at the rasterization stage by prioritizing Gaussians with high contributions while reducing computations for those with low contributions.
Both techniques can be seamlessly integrated into existing 3DGS pipelines with minimal fine-tuning.
Collectively, our framework achieves up to 6.3$\times$ speedup and 39.1\% model reduction while achieving superior rendering quality compared to existing methods.
Our codes will be released upon publication.

\end{abstract}    
\section{Introduction}
\label{sec:intro}

3D Gaussian Splatting (3DGS)~\cite{kerbl20233d} has emerged as a vital technique in many rendering-related domains, including autonomous driving~\cite{liu2025citygaussian, liu2024citygaussianv2, kerbl2024hierarchical, tancik2022block}, augmented and virtual reality (AR/VR)~\cite{rojas2023re, chen2023mobilenerf, hu2022efficientnerf, hedman2021baking}. 
These real-time applications require high rendering performance to ensure seamless quality of services.
However, the limited computational resources of mobile platforms often constrain these applications, preventing them from reaching their full potential.

For instance, Nvidia AGX Orin, a leading-edge compute module for modern vehicles, offers only 3.4\% of the computational resources of a Nvidia A100 workstation GPU. 
Similarly, the Snapdragon XR2 chip, the most widely used chip for AR/VR devices, has 4.2\% of the computational power compared to Nvidia RTX 4090. 
On Nvidia’s AGX Orin, 3DGS~\cite{kerbl20233d} barely achieves 20 frames-per-second (FPS) on real-world datasets~\cite{barron2021mip, knapitsch2017, hedman2018deep}, far from the real-time requirement, i.e., 90 FPS of VR~\cite{wang2023effect}.
An interesting challenge is to achieve real-time on devices with only 3-4\% of the resources compared to high-end GPUs.

To achieve this goal, we must understand the major performance bottlenecks in the current 3DGS rendering pipelines.
We summarized the bottlenecks into three aspects: computational intensity, rendering inefficiency, and memory budget. 
This paper proposes a \textit{one-stop solution} framework, \proj, that addresses the inefficiencies across these three dimensions (\Sect{sec:method}).
Next, we briefly discuss the inefficiencies in today’s 3DGS rendering pipeline and highlight how our proposed solution addresses these limitations.

\begin{figure}
    \centering
    \includegraphics[width=0.95\columnwidth]{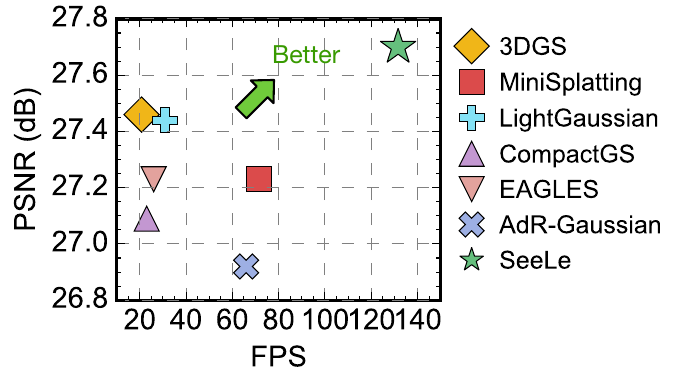}
    \caption{Our acceleration framework, \proj, achieves up to 6.3$\times$ speedup against the state-of-the-art 3DGS algorithms.}
    \label{fig:teaser}
\end{figure}

\paragraph{Computational Intensity.}
One primary challenge in the 3DGS pipeline is its computational intensity. 
From an individual pixel's perspective, rendering one pixel typically involves processing thousands of Gaussian points, with each GPU thread rendering one pixel. 
% ~\footnote{We use ``Gaussians'' and ``Gaussian points'' interchangeably: there is a one-to-one mapping between them.}
This rendering process imposes substantial computational overhead on resource-constrained devices. 
Prior solutions~\cite{fan2023lightgaussian, fang2024mini} primarily focus on reducing the overall model size, namely the number of Gaussian points, but these methods often trade off rendering quality for performance.

In contrast, we propose \textit{hybrid preprocessing} with a view-dependent scene representation that dynamically loads the relevant Gaussians to GPU memory during runtime (\Sect{sec:method:hp}). 
Our data representation naturally identifies the necessary Gaussians contributing to the current viewpoint while leaving other irrelevant Gaussians untouched.
Meanwhile, our method does not make any compromise between performance and accuracy. 
On average, this method alone outperforms the state-of-the-art rendering algorithms while achieving a 2.8$\times$ speedup.

\paragraph{Rendering Inefficiency.}
The second issue with the current 3DGS rendering pipeline is that all splatted Gaussians go through the same rasterization pipeline regardless of their contribution to the final pixels. 
Our experiment shows that, for each pixel, 1.5\% of the Gaussians contribute to 99\% of the final pixel.
Therefore, we argue that this uniform treatment to all Gaussians leads to severe rendering inefficiencies, as \textit{less significant Gaussians should naturally receive fewer computational budgets}.
However, prior work~\cite{lee2024compact, girish2025eagles, lin2024rtgs}, such as pruning or quantization techniques, still retains a uniform rendering pipeline and fails to differentiate Gaussians based on their contributions. 

On the contrary, we propose contribution-aware rasterization from a pixel-centric perspective (\Sect{sec:method:cr}). 
Rather than reducing computation uniformly, we dynamically identify and prioritize Gaussians with high contributions (i.e., high accumulative transparencies) and reduce computation budgets for those with low contributions. 
Gaussians with low contributions, which typically contribute to low-frequency textures (as shown in \Fig{fig:point_contribution}), are allocated less computation by dynamically skipping insignificant operations. 
We show that this method achieves an additional 1.4$\times$ speedup while retaining the same rendering quality.

\paragraph{Memory Budget.}
Our last contribution eases the peak GPU memory.
As model complexity scales up, the GPU memory requirements often become unsustainable due to the increasing number of Gaussians involved. 
Solutions~\cite{lee2024compact, ren2024octree} apply offline compression techniques to reduce memory usage, but they overlook a key opportunity: rendering a given viewpoint does not require loading all Gaussians into GPU memory simultaneously.

Leveraging this key observation, we design proactive memory management that asynchronously loads necessary data to GPU memory without disrupting the critical path of rendering. 
Combined with our view-dependent scene representation, our memory management significantly reduces peak GPU memory usage. 
On average, our approach achieves a 39.1\% reduction in model size, allowing large-scale rendering on mobile devices.

Collectively, our framework achieves up to 6.3$\times$ speedup and 39.1\% runtime model reductions, all while retaining a better rendering quality compared to state-of-the-art algorithms. 
Furthermore, our techniques are orthogonal to existing optimizations and can be seamlessly integrated with other mainstream Gaussian splatting pipelines.

The contribution of this paper is summarized as follows:
\begin{itemize}
    % \vspace{-5pt}
    \item We introduce a view-dependent scene representation that combines online and offline filtering to reduce computation overhead and runtime GPU memory.
    % \vspace{-5pt}
    \item We propose a contribution-aware rasterization algorithm that dynamically skips the insignificant computations and improves the parallel efficiency.
    % \vspace{-5pt}
    \item We design an integrated co-training procedure that integrates the aforementioned techniques and achieves better rendering quality against the corresponding baselines.
\end{itemize}

\section{Related Work}
\label{sec:related}

\subsection{Efficient Data Representation}

A 3DGS model is essentially a set of points in 3D space that represent the physical structure of the world. 
The model size often becomes one of the key bottlenecks in rendering performance and memory usage.
To reduce the model size, several studies have proposed various data representations to optimize storage and performance.
Some~\cite{niedermayr2024compressed, chen2024hac, chen2025hac++, papantonakis2024reducing, durvasula2025contrags, niemeyer2024radsplat} reduce GPU memory usage by leveraging encoding techniques to compress the model. 
For example, CompactGS~\cite{lee2024compact} employs vector quantization to compress the model offline and decodes it at runtime using a pretrained codebook. Similarly, EAGLES~\cite{girish2025eagles} uses an encoding network to compress the neural radiance field into a compact representation.
However, both methods introduce non-trivial execution overhead during runtime, which limits their applicability in computation-constrained scenarios.

Other pruning-based approaches use a learnable mask~\cite{liu2025maskgaussian, zhang2024lp, taghipour2025svr} or a significance score~\cite{fang2024mini, fang2024mini2, mallick2024taming, fan2023lightgaussian, hanson2025pup} to eliminate redundant Gaussians. However, these methods have to make a trade-off between rendering quality and efficiency.
Some studies adopt structured data representations such as octrees~\cite{ren2024octree} and kd-tree~\cite{kerbl2024hierarchical} to organize Gaussian points while some researches~\cite{hamdi2024ges, held20253d, li20253d, ye2025gaussian} explore alternative primitives for more efficient scene representation. 
However, they introduce additional runtime overhead in loading or rasterizing Gaussians.

In contrast to prior studies, we focus on real-time performance.
Our view-dependent scene representation achieves significant model size reduction with minimal runtime overhead and no quality compromise.

\subsection{Rasterization Optimization}

In addition to pruning Gaussian points, another line of research focuses on optimizing the performance of the 3DGS pipeline itself.
Recent studies propose various online filtering techniques, such as axis-aligned bounding box (AABB)~\cite{klosowski1998efficient, wang2024adr, hanson2025speedy} and oriented bounding box (OBB)~\cite{gottschalk1996obbtree, lee2024gscore} intersection tests, to perform more fine-grained filtering before rasterization. 
These works aim to reduce the number of Gaussians processed during the rasterization stage at runtime, thereby reducing the computational workload.

On the other hand, some methods~\cite{feng2025flashgs, liao2025tc, gui2024balanced} focus on optimizing the rasterization stage. 
For example, FlashGS~\cite{feng2025flashgs} designs a software pipelining to overlap data fetching with rasterization, improving overall efficiency. 
In contrast, Balanced3DGS~\cite{gui2024balanced} addresses workload imbalance by introducing an offline scheduling mechanism to rebalance workloads at both the block and tile levels.
Additionally, a few studies~\cite{lee2024gscore, lin2025metasapiens, pei2025gcc, lee2025vr} explore the new hardware designs to support 3DGS, further advancing the efficiency and scalability of the pipeline.

Unlike prior works, our optimization takes a unique angle to accelerate 3DGS by addressing the inherent contribution imbalance across Gaussian points. 
Our method dynamically assigns computational budgets to Gaussians based on their contributions to the final rendered image. 
Moreover, our algorithm is naturally GPU-friendly and alleviates execution inefficiencies such as warp divergence, achieving better overall performance with no hardware modifications.

\section{Methodology}
\label{sec:method}

\begin{figure*}
    \centering
    \includegraphics[width=\textwidth]{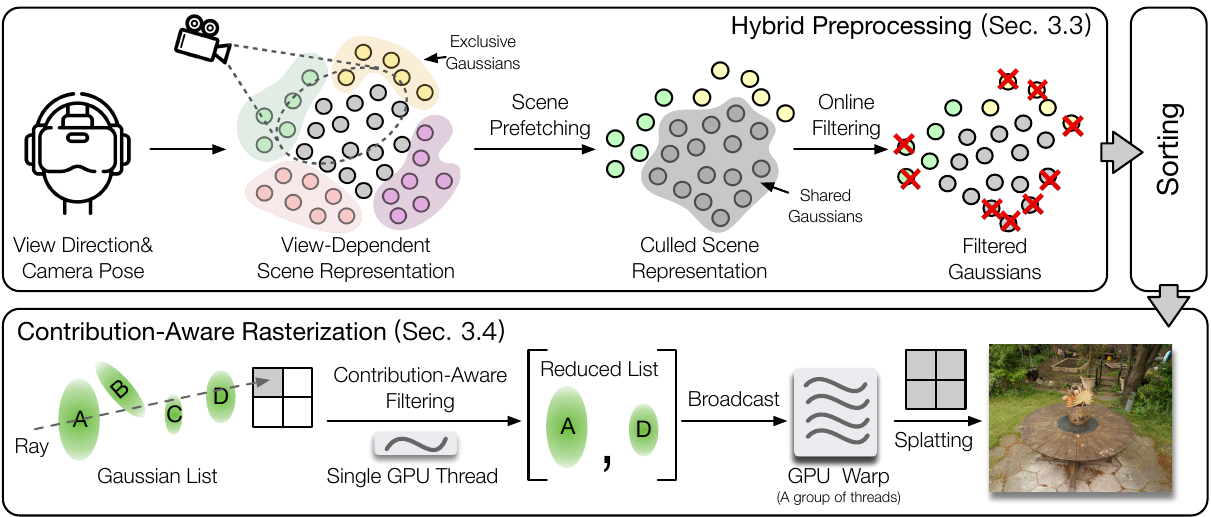}
    \caption{The overview of \proj. we modify the two steps, \textit{preprocessing} and \textit{rasterization}, and propose two novel techniques: \textit{hybrid preprocessing} and \textit{contribution-aware rasterization}, in Gaussian splatting. Hybrid preprocessing leverages offline coarse-grained scene clustering and online filtering to reduce the number of Gaussians before rasterization. Contribution-aware rasterization dynamically identifies insignificant Gaussians and skips them to accelerate the overall rendering pipeline.}
    \label{fig:overview}
\end{figure*}

\subsection{Preliminaries}
\label{sec:method:pre}

3DGS algorithms adhere to an explicit point-based representation, where each point is modeled by a Gaussian ellipsoid (a.k.a., Gaussian for short) that captures the geometric and textural properties of the scene.
The geometric property of a Gaussian is governed by its 3D centroid position $x$ and a 3D covariance matrix $\Sigma$.
The textural properties are described by two components: its opacity $o$ and spherical harmonic (SH) coefficients.
Using these properties, the 3DGS pipeline renders a frame on a tile-by-tile basis via three steps: \textit{preprocessing}, \textit{sorting}, and \textit{rasterization}.

\paragraph{Preprocessing.}
Given a camera pose and a view direction, the preprocessing stage first filters out Gaussians outside the current view frustum.
Meanwhile, this stage identifies the intersection between Gaussians and rendering tiles.

\paragraph{Sorting.} Once each tile collects its intersected Gaussians, the sorting stage determines the rendering order of those Gaussians.
This stage ensures that all points are rendered, from the closest to the furthest, based on their depth. 

\paragraph{Rasterization.} Once all Gaussians are sorted, the rasterization stage renders these Gaussians tile-by-tile. 
Within each tile, every pixel iterates through the same set of Gaussians, computing transparency and accumulating their colors into the pixel in the sorted order. 
\Eqn{eqn:nerf} governs the color accumulation process of pixel $\textbf{p}$,
\begin{align}
\label{eqn:nerf}
   C(\textbf{p}) & = \sum_{i=1}^{N} \Gamma_i \alpha_i \textbf{c}_{i},\ \text{where} \ \Gamma_i = \prod^{i-1}_{j=1} (1-\alpha_j),
\end{align}
where $\Gamma_i$ denotes the accumulative transmittance of pixel $\textbf{p}$ from the Gaussian $1$st to the $i-1$th. $\alpha_i$ and $\textbf{c}_i$ stand for the transparency and the color at the $i$th Gaussian, respectively.
The transparency of a Gaussian $\alpha_i$ is calculated by the Gaussian opacity $o_i$, the 2D position $x'$ and the 2D covariance matrix, $\Sigma'$~\cite{zwicker2001ewa},
\begin{equation}
\label{eqn:alpha}
    \alpha_i=o_i e^{-\frac{1}{2}(\mathbf{p}-x')^T\Sigma'^{-1}(\mathbf{p}-x')}.
\end{equation}
$\textbf{c}_i$ is a function of the Gaussian spherical harmonics.

Note that, if the Gaussian's $\alpha_i$ value is insignificant (less than $\alpha_{\theta}$, i.e., $\frac{1}{255}$), this Gaussian will be skipped in the color accumulation to avoid numerical instabilities. 
The color accumulation process terminates once the cumulative transmittance $\Gamma_i$ surpasses a predefined threshold, $\Gamma_\theta$.

\subsection{Overview}
\label{sec:method:overview}

\Fig{fig:overview} gives the overview of our proposed acceleration framework.
To address three rendering inefficiencies highlighted in \Sect{sec:intro}, we introduce two key components: \textit{hybrid preprocessing} and \textit{contribution-aware rasterization} to reduce the runtime computation and peak memory.

In 3DGS pipelines, rasterization is the most time-consuming stage, accounting for $>$67\% of the execution time on a Nvidia Ampere GPU.
Prior studies have explored two primary methods to reduce the overall workload of rasterization.
The first approach applies pruning techniques, which eliminate insignificant Gaussians offline~\cite{fan2023lightgaussian, fang2024mini}.
The second proposes online filtering techniques, such as AABB and OBB intersection tests, to perform more fine-grained filtering before rasterization~\cite{klosowski1998efficient, wang2024adr, gottschalk1996obbtree, lee2024gscore}.

Our hybrid preprocessing combines the merits of both offline and online techniques.
During the offline processing, we design a view-dependent scene representation that clusters Gaussians based on their contributions to a set of closely related rendering viewpoints.
At runtime, our framework can asynchronously prefetch the relevant clusters into the GPU memory before rendering.
% This approach largely reduces the filtering workload during preprocessing.
% In addition, t
Our view-dependent scene representation inherently eliminates Gaussians that are irrelevant to the current rendering while within the view frustum. 
Combined with online filtering, we further reduce the number of Gaussians intersected for each tile in the subsequent stages.
% our hybrid preprocessing collectively minimizes the processed Gaussians in rasterization.

Meanwhile, we propose a novel contribution-aware rasterization to accelerate the rasterization pipeline itself.
The key idea of our approach is to penalize the computations assigned to insignificant Gaussians purposely, allowing for efficient rendering without compromising quality.

Specifically, our rasterization pipeline organizes pixels into small groups while still keeping one thread responsible for one pixel. 
At runtime, only a single pixel within each pixel group calculates its Gaussian transparency $\alpha$ rather than having all pixels compute $\alpha$. 
The purpose is to let this pixel identify Gaussians that potentially contribute less to the pixel group, so we can exclude these Gaussians from further rasterization. 
% Since the identified Gaussians are likely to have minimal impact on the remaining pixels within a pixel group, we exclude these Gaussians from further rasterization. 
In this way, we eliminate insignificant Gaussian accumulation at runtime. 
Combined with our integrated co-training, we can achieve a smoother rendering experience.
A more rigorous explanation of this approach is provided in \Sect{sec:method:cr}.

\subsection{Hybrid Preprocessing, $\cHP$}
\label{sec:method:hp}

% \begin{figure}
%     \centering
%     \includegraphics[width=\linewidth]{cluster_representation}
%     \caption{Our scene representation clusters all Gaussians into shared Gaussians and exclusive Gaussians. Here, we show the Gaussian positions \textit{without} scales. The yellow points in \Fig{fig:representation}(b) represent the shared Gaussians, while the remaining colored points correspond to the exclusive Gaussians in different clusters.}
%     \label{fig:representation}
% \end{figure}

\begin{figure}
  \centering
  \begin{subfigure}{0.48\linewidth}
    \includegraphics[width=\columnwidth]{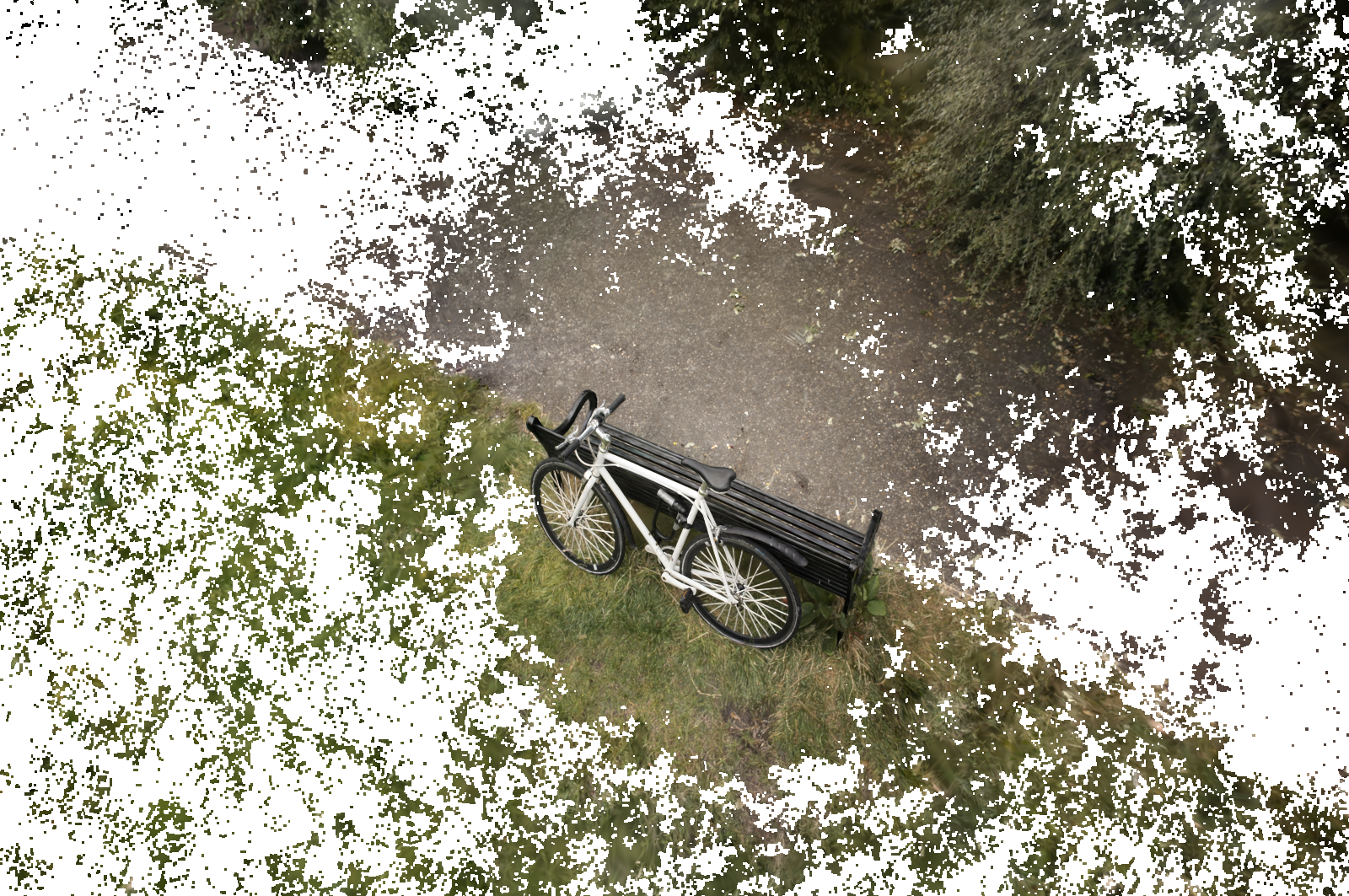}
    \caption{Original Gaussian points.}
    \label{fig:points}
  \end{subfigure}
  \hfill
  \begin{subfigure}{0.48\linewidth}
    \includegraphics[width=\columnwidth]{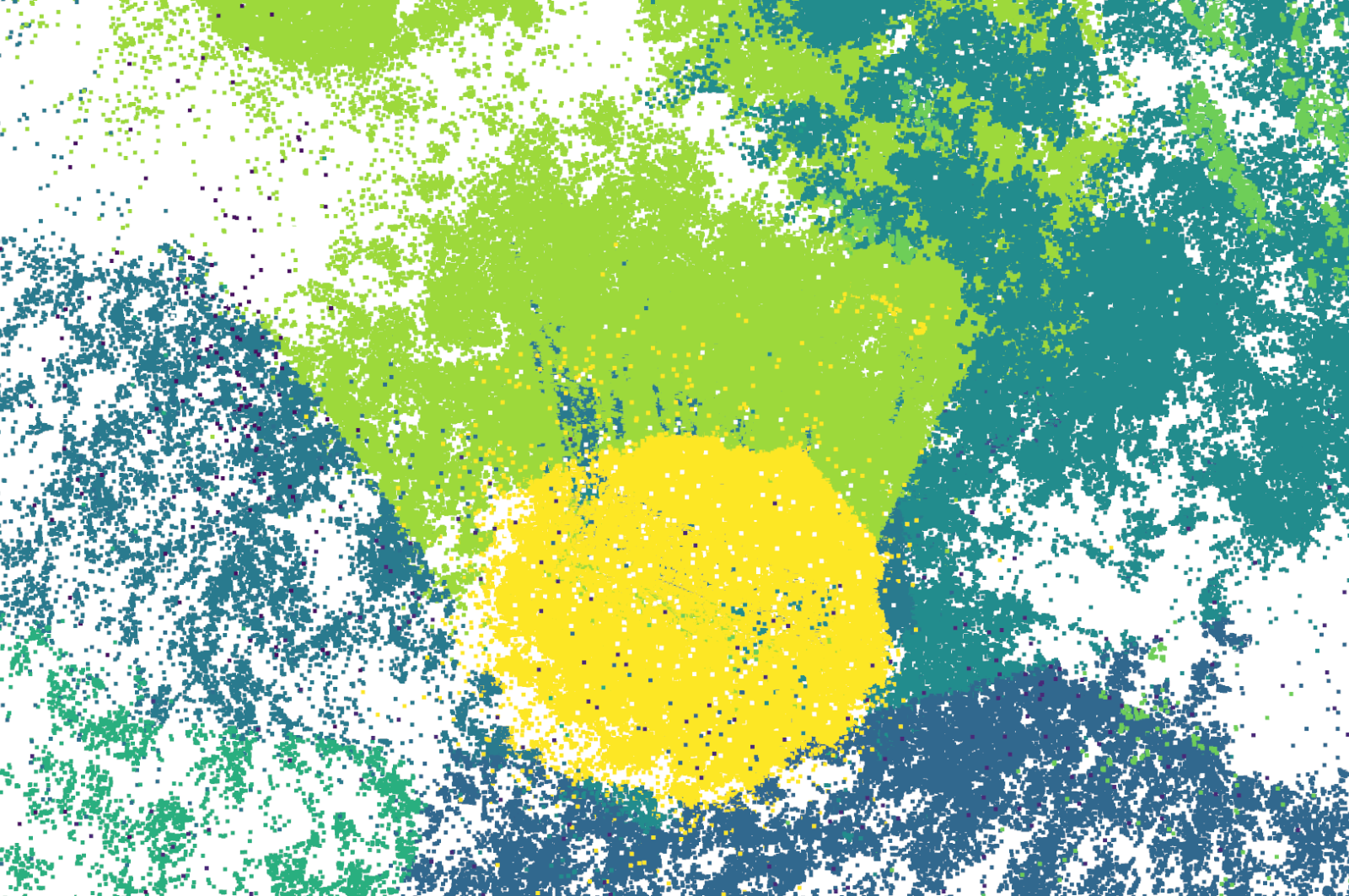}
    \caption{Clusters w/ shared points.}
    \label{fig:cluster_w_shared}
  \end{subfigure}
  \caption{Our scene representation clusters Gaussians into shared ones and exclusive ones. Here, we show the Gaussian positions \textit{without} scales. The yellow points in \Fig{fig:cluster_w_shared} represent the shared Gaussians, while the other colors correspond to the exclusive Gaussians in different clusters.}
  \label{fig:representation}
\end{figure}

\paragraph{View-Dependent Scene Representation.}
The idea of this data representation is to preprocess Gaussians offline, thereby reducing computational effort during the online preprocessing stage.
Our key observation is that different viewpoints rely on distinct sets of Gaussians for rendering. 
Viewpoints in close proximity are likely to share similar Gaussians, while those farther apart share fewer. 

Based on this observation, our data representation classifies Gaussians into two categories: shared Gaussians and exclusive Gaussians, as shown in \Fig{fig:representation}.
Shared Gaussians are shared across all viewpoints and are frequently used for rendering frames from any viewpoint (yellow points in \Fig{fig:cluster_w_shared}).
On the other hand, exclusive Gaussians are specifically attached to a group of viewpoints and are only used for rendering frames within that group (other colors in \Fig{fig:cluster_w_shared}).
% For instance, \Fig{fig:points} illustrates the original set of Gaussian points, which are then clustered as shown in \Fig{fig:cluster_w_shared}. 
% The yellow points in \Fig{fig:cluster_w_shared} represent the shared Gaussians, while the remaining colored points correspond to the exclusive Gaussians in different clusters.

\paragraph{Offline Clustering.} The key to our approach lies in how to classify the Gaussians into shared and exclusive Gaussians.
Initially, our method randomly samples camera poses and clusters them into $N$ smaller groups based on the weighted similarity of their camera positions, $\overrightarrow{x}$ and view directions, $\overrightarrow{v}$. 
Specifically, we first normalize the camera positions, $\overrightarrow{x}$, then concatenate normalized $\overrightarrow{x}$ and $\overrightarrow{v}$ as a 1D dimensional vector, $(\overrightarrow{x}, \overrightarrow{v})$, where $\beta$ is a hyperparameter that balances the impact of position and orientation. 
Here, $\beta$ is set to 1 for clustering all Gaussians.
% $(\overrightarrow{x}, \beta \overrightarrow{v})$ is used to cluster all Gaussians.
% Specifically, we first normalize the camera positions, $\overrightarrow{x}$, then concatenate normalized $\overrightarrow{x}$ and $\overrightarrow{v}$ as a 1D dimensional vector, $(\overrightarrow{x}, \beta \overrightarrow{v})$, where $\beta$ is a hyperparameter that balances the impact of position and orientation. $(\overrightarrow{x}, \beta \overrightarrow{v})$ is used to cluster all Gaussians.

For each cluster, we identify and reserve the Gaussians that are the primary contributors to pixel rendering.
Specifically, for each pixel, we find the Top-$k$ Gaussians based on their contribution to the cumulative transmittance, $\Gamma_i \alpha_i$, as the contribution metric. 
The rationale here is that the remaining Gaussians contribute minimally to the final pixel and can be safely discarded without compromising rendering quality.
We choose a $k$ value of 32, as we find that increasing $k$ further yields minimal quality improvement.
% \fixme{We discuss the sensitivity of rendering quality to the Top-$k$ value in \Sect{sec:eval:sens}.}

Once we identify the top Gaussian contributors for each cluster, we find the union of top contributors across all clusters and grant these top Gaussians to be shared across all clusters.
These granted Gaussians are then the shared Gaussians in our representation. 
The remaining Gaussians within each cluster are classified as exclusive Gaussians.

\paragraph{Algorithm.}
During the actual rendering, for a given camera pose, our approach first identifies the nearest cluster based on the cluster centroid. 
Next, all Gaussians belonging to this cluster and its $M$ neighboring clusters are selected, as shown in \Fig{fig:overview}.
In this example, green points are the nearest cluster, and the yellow points are the nearest neighbors to that cluster ($M$ is set to 1 in this toy example).
The selected $1+M$ clusters are used to render the current camera pose instead of all Gaussians.
Note that, shared Gaussians, gray points in \Fig{fig:overview}, are always stored in GPU memory since they are used for any poses.
Unlike the original 3DGS, our approach retains only the relevant clusters to stay in GPU memory, thus reducing the peak GPU memory usage.
Empirically, we set $M$ to be 3, which strikes a good balance between performance and rendering quality.

\paragraph{Online Filtering.}
Once these Gaussians are loaded, they are transformed into screen space with a new 2D Gaussian position $x'$ and covariance matrix $\Sigma'$. 
To filter unnecessary Gaussians for each tile, the intersection test typically uses $3\sigma$ envelope to approximate,
\begin{equation}
\sqrt{(\mathbf{p} - x_i')^T \Sigma_i'^{-1}(\mathbf{p} - x_i')} = 3,
\end{equation}
where $\mathbf{p}$ is the pixel position.
Due to the redundancy of this approximation~\cite{hanson2025speedy, pei2025gcc}, we further include the opacity $o_i$ into our online filtering equation as follows. 
As we would skip insignificant $\alpha_i$, which value is less than $\alpha_{\theta}$, $\frac{1}{255}$, \Eqn{eqn:alpha} can be expressed as,
\begin{equation}
\sqrt{(\mathbf{p} - x_i')^T \Sigma_i'^{-1}(\mathbf{p} - x_i')} = \sqrt{2\ln\frac{o_i}{\alpha_{\theta}}}.
\end{equation}
Therefore, we execute fine-grained online filtering by including opacity into consideration,
\begin{equation}
(\mathbf{p} - x_i')^T \Sigma_i'^{-1}(\mathbf{p} - x_i') = \min(2\ln\frac{o_i}{\alpha_{\theta}}, 9).
\end{equation}
This further eliminates false-positive intersected Gaussians when performing Gaussian-tile intersections.
% The red crosses in \Fig{fig:overview} highlight the false-positive Gaussians.
% \Sect{sec:eval:ab} shows the effectiveness of our hybrid preprocessing.

\paragraph{Scene Prefetching.}
Our hybrid preprocessing not only enhances rendering performance but also reduces peak GPU memory usage. 
However, loading Gaussian clusters at runtime could introduce non-trivial execution overhead and potentially cause frame stuttering.
To address this, we prefetch future clusters ahead of time by asynchronously loading them using a dedicated GPU stream. We adopt similar prediction methods as prior works~\cite{hou2020motion, feng2024cicero} and find that linearly extrapolating the camera pose is an effective method for predicting future poses in our scene prefetching.
This way, our approach enables asynchronous prefetching to be overlapped with runtime frame rendering. 
Compared to loading all Gaussian clusters into GPU memory, our scene prefetching reduces the runtime model size by 39.1\% with a minimal runtime overhead ($<$6\% of the total latency).

\subsection{Contribution-Aware Rasterization, $\cCR$}
\label{sec:method:cr}

\begin{figure}
    \centering
    \includegraphics[width=\columnwidth]{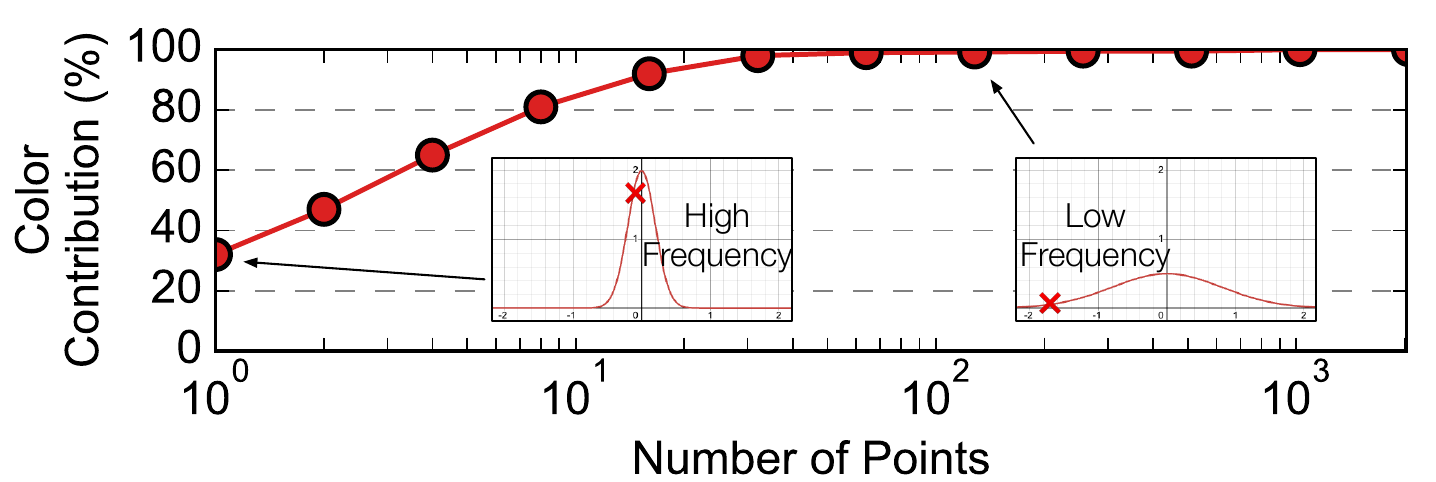}
    \caption{The significance of Gaussians towards the final pixel. Gaussians are sorted in descending order. We empirically find that significant Gaussians are typically sampled in high-frequency regions, while insignificant Gaussians are more likely to be sampled in low-frequency regions (red crosses).}
    \label{fig:point_contribution}
\end{figure}

\paragraph{Motivation.}
The goal of our contribution-aware rasterization is to pay ``computational'' attention to significant Gaussians that contribute more to the final pixel values while reducing computations for less important Gaussians. 
Our observation is that the color values of most pixels are determined by a small fraction of Gaussians, as illustrated in \Fig{fig:point_contribution}. 
We characterize the MipNeRF360 dataset~\cite{barron2021mip} and draw the cumulative transparency of each pixel by sorting the Gaussian transparency contribution $\Gamma \alpha$ in descending order.
The result in \Fig{fig:point_contribution} shows that 1.5\% of the top Gaussians contribute to 99\% of the final pixel, while remaining Gaussians contribute minimally due to the lower sampling $\alpha$.
% The remaining Gaussians contribute minimally, primarily adding low-frequency textures.
Furthermore, these low contribution samples typically occur in Gaussians with larger variances or at locations farther from the mean, corresponding to the low-frequency components and sharing low gradients and good spatial locality. 
More rigorous analysis of the correlation between Gaussians and frequency is shown in the supplementary.
% This is not an artifact but an inherent feature of Gaussian representation. 
% Assuming uniform initial opacity across all Gaussians, low $\alpha$ values tend to encode broad and low-frequency features, whereas high $\alpha$ values (with high gradients) are often associated with high-frequency details, 

\begin{algorithm}[t]
\caption{Contribution-Aware Rasterization}\label{algo:raster}
\KwData{a list of Gaussians $ \cG $, a group of pixels $ \cP $ }
\KwResult{an image tile $ \cI $}
$\cI.\text{init()}$\;
\For{$g \in \cG$}{
    $\alpha_0 \leftarrow \text{ComputeTransparency}(\cP[0], g)$;
    
    \If{ $\alpha_0 < \alpha_{\theta}$  }{
        continue;
    }
    
    \For{ $p \in \cP$ }{
        $\alpha_{p,g} \leftarrow \text{ComputeTransparency}(p, g)$;
        
        $\cI \leftarrow \text{AccumulateColor}(\alpha_{p,g}, p)$;
    }
    \text{CheckTermination()};
}
\end{algorithm}

\paragraph{Algorithm.}
To leverage the observation, we propose a \textit{contribution-aware rasterization} algorithm to identify insignificant samples and rearrange computational resources, as outlined in \Alg{algo:raster}.
Instead of uniformly assigning the same computation to all Gaussians, our algorithm organizes every $w \times w$ pixels into a small group, $\cP$. 
During each iteration, only the leader pixel $p$ within a group, which is the centroid of this group, computes the $\alpha$ value for a given Gaussian, $g$. 
If this $\alpha$ value falls below a predefined threshold $\alpha_{\theta}$ ($\frac{1}{255}$, to avoid numerical instability), the algorithm skips the color blending of this Gaussian on all pixels within $\cP$. 
We set $w$ to 2 to balance rendering efficiency and quality, and analyze the upper bound on the blending error introduced by this skipping in the appendix. 
% TODO: delete param config in sec 4.1?

% This skipping reduces the workload for insignificant Gaussians. 

For Gaussians deemed important (i.e., those with $\alpha$ values above the threshold, $\alpha_{\theta}$), our algorithm performs the same rasterization computations as the original approach, ensuring rendering quality is preserved. 
This dynamic computation reassignment improves rendering efficiency without compromising visual quality.

\begin{figure}
    \centering
    \includegraphics[width=\columnwidth]{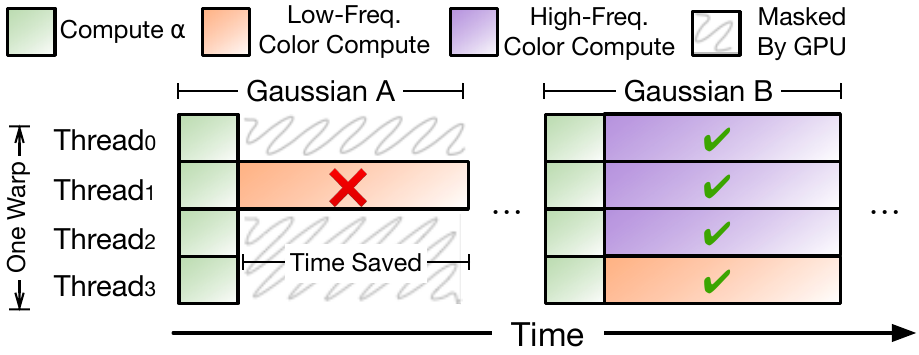}
    \caption{An example of warp divergence in GPU. All threads compute $\alpha$ and then perform color blending in ``lockstep''. Our algorithm can detect the insignificant Gaussians (e.g., Gaussian $A$) and skip their color blending, as highlighted by the red cross. Thus, we save the total execution time.}
    \label{fig:warp}
\end{figure}

\paragraph{GPU Parallelism.}
Not only does our algorithm reduce the overall computation, but it also inherently achieves higher parallelism by alleviating warp divergence.
In the GPU implementation, pixels within a group are assigned to one warp.
All pixels within a warp execute in ``lockstep'' as shown in \Fig{fig:warp}.
For instance, when blending Gaussian $A$, only thread $1$ accumulates the insignificant Gaussian $A$ into its pixel, while other threads wait for thread $1$ to complete.

In contrast, our algorithm only lets thread $0$ detect if Gaussian $A$ is insignificant.
In this case, all threads in this warp skip the color blending of Gaussian $A$.
In the case of Gaussian $B$, if it is considered significant, all threads would blend Gaussian $B$.
This ensures that these pixels either skip or process the same Gaussians collectively. 
This way, we eliminate divergence within a warp.

\subsection{Integrated Fine-Tuning}
\label{sec:method:ct}

Note that, both $\cHP$ and $\cCR$ modify the rendering pipeline and might impact the view consistency.
To maintain a better view consistency across frames, our algorithm only requires minimal fine-tuning to achieve better rendering quality.
Since both \textit{hybrid preprocessing} and \textit{contribution-aware rasterization} do not participate in gradient descent, these two techniques can be integrated into the current training process without any modifications to back-propagation.

Once the 3DGS model is trained by the canonical training process, we convert the model into our view-dependent scene representation in \Sect{sec:method:hp}.
% Next, we fix the shared Gaussians and fine-tune only the exclusive Gaussians in each cluster individually for 1000 steps.
Next, we fine-tune the model for additional 1000 steps with our loss function,
\begin{equation}
    \cL_{total} = \cL_{\text{3DGS}} + \gamma * \cL_{\text{consistency}},
\end{equation}
where $\cL_{\text{3DGS}}$ is the original loss function in 3DGS and $\cL_{\text{consistency}}$ is the view consistency loss. Here, we use \reflectbox{F}LIP score~\cite{andersson2020flip} with 7 frames as our metric. $\gamma$ is set to be 0.1.

% Since our algorithm always takes $1+M$ clusters (1 nearest cluster and $M$ neighboring clusters) during rendering, it naturally accommodates the view discrepancies when shifting between two cluster boundaries.
% Lastly, we fix the exclusive Gaussians and fine-tune the shared ones for 1000 steps.

\begin{table*} 
\caption{Quantitative evaluation of our method against the state-of-the-arts~\cite{kerbl20233d, fang2024mini, fan2023lightgaussian, wang2024adr}. The \green{bold green} results highlight the better results between ours, \proj, and the corresponding baselines. 
\proj achieves better quality across all three quality metrics with an average 2.4$\times$ speedup. 
\gold and \silver denote the \textit{best} and \textit{second-best} results among all methods, respectively.}
\resizebox{\textwidth}{!}{
\renewcommand*{\arraystretch}{1}
\renewcommand*{\tabcolsep}{3pt}
\begin{tabular}{ c|rrrrrr|rrrrrr|rrrrrr } 
\toprule[0.15em]
Dataset & \multicolumn{6}{c|}{\textbf{Mip-NeRF360}} & \multicolumn{6}{c|}{\textbf{Tanks\&Temples}} & \multicolumn{6}{c}{\textbf{Deep Blending}} \\ 
\midrule[0.05em]
\multirow{2}{*}{Metrics} & \multicolumn{3}{c|}{Quality} & \multicolumn{3}{c|}{Efficiency} & \multicolumn{3}{c|}{Quality} & \multicolumn{3}{c|}{Efficiency} & \multicolumn{3}{c|}{Quality} & \multicolumn{3}{c}{Efficiency} \\ 
& PSNR$\uparrow$ & SSIM$\uparrow$ &  \multicolumn{1}{r|}{LPIPS$\downarrow$}  & \multicolumn{1}{c}{FPS$\uparrow$} & \specialcell{\#Inst.\\($10^6$)$\downarrow$} & \specialcell{Mem.\\(MB)$\downarrow$} & PSNR$\uparrow$ & SSIM$\uparrow$ & \multicolumn{1}{r|}{LPIPS$\downarrow$} & \multicolumn{1}{c}{FPS$\uparrow$} & \specialcell{\#Inst.\\($10^6$)$\downarrow$} & \specialcell{Mem.\\(MB)$\downarrow$} & PSNR$\uparrow$ & SSIM$\uparrow$ & \multicolumn{1}{r|}{LPIPS$\downarrow$} & \multicolumn{1}{c}{FPS$\uparrow$} & \specialcell{\#Inst.\\($10^6$)$\downarrow$} & \specialcell{Mem.\\(MB)$\downarrow$} \\ 
\midrule[0.05em]
% CompactGS~\cite{lee2024compact} & 26.99 & 0.797 & 0.247 & 24.18 & 1824.57 & 122.7\ \ \ & 23.42 & 0.831 & 0.202 & 46.53 & 920.04 & 93.0\ \ \  & 29.84 & 0.902 & 0.257 & 32.87 & 1294.45 & 104.0\ \ \  \\
% EAGLES~\cite{girish2024eagles} & 27.14 & 0.805 & 0.239 & 22.30 & 1988.56 & 68.1\ \ \  & 23.34 & 0.836 & 0.200 & 60.27 & 713.29 & 34.0\ \ \  & 29.95 & 0.907 & 0.249 & 25.24 & 1555.53 & 61.5\ \ \  \\

3DGS~\cite{kerbl20233d} & 27.46 & 0.812 & 0.223 & 20.79 & 2168.68 & 710.6\ \ \  & 23.75 & \silver0.848 & 0.176 & 41.97 & 1034.37 & 430.9\ \ \  & 29.59 & 0.902 & 0.244 & 23.94 & 1645.11 & 639.9\ \ \  \\
\proj+ 3DGS & \gold\green{27.72} & \silver\green{0.814} & \silver\green{0.217} & \green{59.67} & \green{778.25} & \green{380.9}\ \ \  & \gold\green{24.02} & \gold\green{0.852} & \gold\green{0.168} & \green{127.80} & \green{356.15} & \green{207.6}\ \ \  & \green{29.79} & \green{0.903} & \gold\green{0.240} & \green{86.58} & \green{421.74} & \green{400.6}\ \ \  \\
\midrule[0.05em]
MiniSplatting~\cite{fang2024mini} & 27.23 & 0.814 & 0.222 & 71.31 & 797.96 & 145.7\ \ \  & 23.18 & 0.837 & 0.187 & 143.27 & 329.69 & 83.3\ \ \  & \gold\green{30.04} & \gold\green{0.908} & 0.243 & 120.01 & 382.86 & 154.9\ \ \  \\
\proj+ MiniSplatting & \silver\green{27.70} & \gold\green{0.822} & \gold\green{0.212} & \gold\green{131.62} & \silver\green{436.41} & \green{106.5}\ \ \  & \green{23.74} & \green{0.846} & \silver\green{0.179} & \gold\green{268.03} & \silver\green{172.40} & \green{60.1}\ \ \  & \silver{30.02} & \silver\green{0.908} & \silver\green{0.242} & \gold\green{200.62} & \silver\green{224.58} & \green{129.6}\ \ \  \\
\midrule[0.05em]
LightGaussian~\cite{fan2023lightgaussian} & 27.44 & 0.807 & 0.235 & 30.89 & 1533.50 & \silver 59.4\ \ \  & 23.82 & 0.842 & 0.189 & 65.60 & 699.71 & \silver 33.8\ \ \  & \green{29.74} & \green{0.901} & 0.250 & 43.17 & 1097.82 & \silver 52.4\ \ \  \\
\proj+ LightGaussian & \green{27.56} & \green{0.810} & \green{0.229} & \green{76.36} & \green{589.85} & \gold\green{39.1}\ \ \  & \silver\green{23.91} & \green{0.846} & \green{0.181} & \green{183.86} & \green{259.78} & \gold\green{20.7}\ \ \  & 29.73 & 0.900 & \green{0.249} & \green{131.16} & \green{324.25} & \gold\green{41.4}\ \ \  \\
\midrule[0.05em]
AdR-Gaussian~\cite{wang2024adr} & 26.92 & 0.786 & 0.267 & 65.95 & 885.27 & 288.6\ \ \  & 23.42 & 0.835 & 0.202 & 114.18 & 442.99 & 203.0\ \ \  & \green{29.68} & \green{0.901} & 0.255 & 78.87 & 485.00 & 330.5\ \ \  \\
\proj+ AdR-Gaussian & \green{27.19} & \green{0.790} & \green{0.261} & \silver\green{96.25} & \gold\green{327.26} & \green{147.6}\ \ \  & \green{23.78} & \green{0.838} & \green{0.196} & \silver\green{193.83} & \gold\green{154.87} & \green{106.2}\ \ \  & 29.64 & 0.899 & \green{0.254} & \silver\green{141.87} & \gold\green{178.16} & \green{127.7}\ \ \  \\
\bottomrule[0.15em]
\end{tabular}
}
\label{tab:overall_eval}
\end{table*}

\section{Evaluation}
\label{sec:eval}

\subsection{Experimental Setup}
\label{sec:eval:exp}

\paragraph{Code Optimization.} 
In addition to the two proposed techniques in \Sect{sec:method}, we also implement the following optimizations.
First, we apply fine-grained data fetching when loading Gaussian attributes to better utilize shared memory on GPU.
Second, we leverage thread-level parallelism to overlap the execution of fetching Gaussian attributes and computation of alpha blending.
Lastly, we replace the compute-intensive \texttt{exp()} function with CUDA intrinsic function \texttt{\_\_exp()}, which generates fewer machine instructions once compiled.
Note that, all optimizations above are nothing more than engineering hacks to improve the program's efficiency; we do not claim them as our major contributions.

\paragraph{Datasets.} 
To show the efficiency and robustness of \proj, we evaluate on three datasets: Mip-NeRF360~\cite{barron2021mip}, Tank\&Temple~\cite{knapitsch2017}, and DeepBlending~\cite{hedman2018deep}.   
For quality evaluation, we use PSNR, SSIM, and LPIPS.
We also report FPS, the number of executed instructions (\#Inst.), and the runtime model size as our performance metrics.

\paragraph{Baselines.} To show the general applicability of our optimization technique, we apply our framework to four state-of-the-art 3DGS algorithms: 3DGS~\cite{kerbl20233d}, MiniSplatting~\cite{fang2024mini}, LightGaussian~\cite{fan2023lightgaussian}, and AdR-Gaussian~\cite{wang2024adr}.

% \paragraph{Training.} By and large, we keep the original training process intact. We train each model for 30000 steps with an additional 2000 fine-tuning steps as described in \Sect{sec:method:ct}.

\paragraph{Algorithm Configurations.}
Unless specified otherwise, we set the number of clusters to be 24 and fix the nearest neighbor cluster, $M$, to be 4 in \Sect{sec:method:hp}. 
The pixel group, $\cP$, is set to $2 \times 2$ in \Sect{sec:method:cr}.
For $2 \times 2$ pixel group, we pick the first pixel as the centroid.

\paragraph{Hardware.}
The primary hardware we used is a mobile Ampere GPU on Nvidia AGX Orin SoC, which is the flagship development board in AR/VR and autonomous driving.
Meanwhile, we also report the performance numbers on Nvidia Orin NX, a low-power embedded computing board, and Nvidia A6000, a powerful workstation GPU.

\subsection{Performance and Quality}
\label{sec:eval:perf}

\begin{table} 
\centering
\caption{The view consistency metrics on Mip-NeRF360.}
\resizebox{\linewidth}{!}{
\renewcommand*{\arraystretch}{1}
\renewcommand*{\tabcolsep}{5pt}
\begin{tabular}{ c|cc|cc|cc|cc } 
\toprule[0.15em]
 & \multicolumn{2}{c|}{\textbf{3DGS}}  & \multicolumn{2}{c|}{\textbf{MiniSplatting}} & \multicolumn{2}{c|}{\textbf{LightGaussian}}  & \multicolumn{2}{c}{\textbf{AdR-Gaussian}}  \\ 
 & \reflectbox{F}LIP$_1$$\downarrow$& \reflectbox{F}LIP$_7$$\downarrow$ & \reflectbox{F}LIP$_1$$\downarrow$  & \reflectbox{F}LIP$_7$$\downarrow$ & \reflectbox{F}LIP$_1$$\downarrow$ & \reflectbox{F}LIP$_7$$\downarrow$ & \reflectbox{F}LIP$_1$$\downarrow$ & \reflectbox{F}LIP$_7$$\downarrow$ \\ 
\midrule[0.05em]
Baseline  & 0.0164 & 0.0466 & 0.0160 & 0.0460 & 0.0168 & 0.0472 & 0.0106 & 0.0332 \\
\proj  & \green{0.0126} & \green{0.0292} & \green{0.0110} & \green{0.0300} & \green{0.0111} & \green{0.0295} & \green{0.0061} & \green{0.0169} \\
\bottomrule[0.15em]
\end{tabular}
}
\label{tab:flip}
\end{table}

% \begin{table} 
% \centering
% \caption{The view consistency metrics, \reflectbox{F}LIP$_1$ and \reflectbox{F}LIP$_7$.}
% \resizebox{\linewidth}{!}{
% \renewcommand*{\arraystretch}{1}
% \renewcommand*{\tabcolsep}{5pt}
% \begin{tabular}{ c|cc|cc|cc } 
% \toprule[0.15em]
%  & \multicolumn{2}{c|}{\textbf{M360 Indoor}}  & \multicolumn{2}{c|}{\textbf{M360 Outdoor}} & \multicolumn{2}{c}{\textbf{M360 Average}}  \\ 
%  & \reflectbox{F}LIP$_1$$\downarrow$& \reflectbox{F}LIP$_7$$\downarrow$ & \reflectbox{F}LIP$_1$$\downarrow$  & \reflectbox{F}LIP$_7$$\downarrow$ & \reflectbox{F}LIP$_1$$\downarrow$ & \reflectbox{F}LIP$_7$$\downarrow$ \\ 
% \midrule[0.05em]
% 3DGS  & \green{0.0031} & 0.0035 & \green{0.0015} & 0.0018 & \green{0.0023} & 0.0027 \\
% \proj+ 3DGS  & 0.0032 & \green{0.0034} & 0.0017 & \green{0.0018} & 0.0024 & \green{0.0026} \\
% \midrule[0.05em]
% MiniSplatting  & \green{0.0018} & \green{0.0020} & 0.0013 & 0.0014  & \green{0.0016} & \green{0.0017} \\
% \proj + MiniSplatting  & 0.0023 & 0.0024 & \green{0.0012} & \green{0.0013} & 0.0018 & 0.0018 \\
% \midrule[0.05em]
% LightGaussian & \green{0.0018} & \green{0.0027} & 0.0016 & 0.0016 & \green{0.0016} & \green{0.0022} \\
% \proj+ LightGaussian  & 0.0021 & 0.0031 & \green{0.0015} & \green{0.0015} & 0.0018  & 0.0023 \\
% \midrule[0.05em]
% AdR-Gaussian & 0.0018 & 0.0027 & 0.0015 & 0.0016 & 0.0016 & 0.0022 \\
% \proj+ AdR-Gaussian  & 0.0021 & 0.0031 & 0.0015 & 0.0015 & 0.0018  & 0.0023 \\
% \bottomrule[0.15em]
% \end{tabular}
% }
% \label{tab:flip}
% \end{table}

\paragraph{Quality Evaluation.}
\Tbl{tab:overall_eval} shows the overall evaluation of \proj on four widely-used 3DGS pipelines: \mode{3DGS}~\cite{kerbl20233d}, \mode{MiniSplatting}~\cite{fang2024mini}, \mode{LightGaussian}~\cite{fan2023lightgaussian}, and \mode{AdR-Gaussian}~\cite{wang2024adr}.
Generally, \proj achieves better rendering quality on all three datasets across all three evaluation metrics.
For instance, on average, \proj improves PSNR and SSIM by 0.28~dB and 0.004, respectively.
More qualitative comparisons are shown in the supplementary.
% We also set up an anonymous website to show the qualitative comparison: \href{https://anonymous-seele.netlify.app/}{link}.

We also measure the view consistency metrics, \reflectbox{F}LIP$_1$ and \reflectbox{F}LIP$_7$~\cite{andersson2020flip} in \Tbl{tab:flip}.
With the loss function proposed in \Sect{sec:method:ct}, \proj achieves better view consistency against the corresponding baselines.

\begin{table} 
\caption{The performance (FPS) over other GPUs: a low-power GPU on Nvidia Orin NX~\cite{orin_nx} and Nvidia A6000~\cite{a6000}.}
\centering
\resizebox{\linewidth}{!}{
\renewcommand*{\arraystretch}{1}
\renewcommand*{\tabcolsep}{3pt}
\begin{tabular}{ c|cc|cc|cc } 
\toprule[0.15em]
Dataset & \multicolumn{2}{c|}{\textbf{Mip-NeRF360}} & \multicolumn{2}{c|}{\textbf{Tanks\&Temples}} & \multicolumn{2}{c}{\textbf{Deep Blending}} \\ 
GPU & Orin-NX & A6000 & Orin-NX & A6000 & Orin-NX & A6000 \\ 
\midrule[0.05em]
3DGS & 3.06 & 134.45 & 6.03 & 197.55 & 3.64 & 155.87 \\
\proj+ 3DGS & \green{9.42} & \green{328.98} & \green{19.63} & \green{536.77} & \green{14.82} & \green{515.60} \\
\midrule[0.05em]
MiniSplatting & 9.62 & 414.02 & 20.57 & 440.34 & 16.84 & 701.51 \\
\proj+ MiniSplatting & \green{17.48} & \green{640.16} & \green{41.78} & \green{1108.65} & \green{32.40} & \green{1031.46} \\
\midrule[0.05em]
LightGaussian & 4.39 & 185.95 & 9.02 & 323.47 & 5.78 & 234.36 \\
\proj+ LightGaussian & \green{12.72} & \green{431.99} & \green{27.45} & \green{755.29} & \green{20.38} & \green{656.60} \\
\midrule[0.05em]
AdR-Gaussian & 10.30 & 383.35 & 16.65 & 456.12 & 13.43 & 478.45 \\
\proj+ AdR-Gaussian & \green{16.70} & \green{524.88} & \green{31.84} & \green{762.71} & \green{19.57} & \green{762.45} \\
\bottomrule[0.15em]
\end{tabular}
}
\label{tab:perf_eval}
\end{table}

\paragraph{Performance Evaluation.}
With comparable rendering quality, \proj consistently achieves speedups across all algorithms.
On average, \proj achieves 3.2$\times$, 1.8$\times$, 2.7$\times$, and 1.7$\times$ speedup on \mode{3DGS}, \mode{MiniSplatting}, \mode{LightGaussian}, and \mode{AdR-Gaussian} respectively. 
The higher speedup observed in \mode{3DGS} compared to the other algorithms can be attributed to the denser nature of \mode{3DGS}.
This density allows \proj to better separate irrelevant Gaussians contributed to different views using \textit{hybrid preprocessing} introduced in \Sect{sec:method:hp}, resulting in greater speedup.
Despite that, sparse models like \mode{MiniSplatting} still benefit from \textit{hybrid preprocessing} due to the view-dependent redundancy, i.e., different views require different Gaussians to render.
\Tbl{tab:overall_eval} also shows the average number of executed instructions of each algorithm using NVIDIA Nsight Compute, \proj indeed reduces the executed instructions, thus overall computation.
\Tbl{tab:abl} further dissects the contributions of our techniques.

\Tbl{tab:perf_eval} shows the speedup of \proj on additional GPUs, Nvidia Orin NX and Nvidia A6000.
The results show that the optimizations proposed in \proj are not tied to a specific GPU architecture, exhibiting good generality across different hardware platforms.
However, we observe that our optimization achieves a higher speedup on lower-power GPUs, as shown in \Tbl{tab:perf_eval}.
% For example, \mode{\proj+3DGS} achieves a $2.9\times$ speedup on the Orin NX, while it achieves a $2.4\times$ speedup on the A6000.
Primarily, low-end GPUs often have more restricted hardware resources.

\proj not only speeds up the overall rendering process but also reduces overall GPU memory consumption.
Here, we exclusively focus on the GPU memory contributed by the Gaussian points, i.e., runtime model weights.
Overall, results show that \proj achieves 39.1\% reduction in runtime model size across three different datasets.
Even for sparse models like \mode{MiniSplatting}, \proj still saves 23.2\% of model size compared to the baselines.
% When considering all other miscellaneous data stored in GPU memory, \proj still achieves the overall 13.19\% GPU memory reduction compared to the original algorithms, on average.

\subsection{Ablation Study}
\label{sec:eval:ab}

\begin{table*} 
\caption{The ablation study of \mode{3DGS} dissects our contributions. \mode{+Opti.} refers to the code optimization in \Sect{sec:eval:exp}, \mode{+HP} represents the \textit{hybrid preprocessing} in \Sect{sec:method:hp}, \mode{+CR} represents the \textit{contribution-aware rasterization} in \Sect{sec:method:cr}.
% , and \mode{\proj} is our full-fledged algorithm.
}
\resizebox{\textwidth}{!}{
\renewcommand*{\arraystretch}{1}
\renewcommand*{\tabcolsep}{3pt}
\begin{tabular}{ c|cccccc|cccccc|cccccc } 
\toprule[0.15em]
Dataset & \multicolumn{6}{c|}{\textbf{Mip-NeRF360}} & \multicolumn{6}{c|}{\textbf{Tanks\&Temples}} & \multicolumn{6}{c}{\textbf{Deep Blending}} \\ 
\midrule[0.05em]
\multirow{2}{*}{Metrics} & \multicolumn{3}{|c|}{Quality} & \multicolumn{3}{c|}{Efficiency} & \multicolumn{3}{c|}{Quality} & \multicolumn{3}{c|}{Efficiency} & \multicolumn{3}{c|}{Quality} & \multicolumn{3}{c}{Efficiency} \\ 
& PSNR$\uparrow$ & SSIM$\uparrow$ &  \multicolumn{1}{c|}{LPIPS$\downarrow$} & FPS$\uparrow$ & \#Inst.($10^6$)$\downarrow$ & Mem.(MB)$\downarrow$ & PSNR$\uparrow$ & SSIM$\uparrow$ & \multicolumn{1}{c|}{LPIPS$\downarrow$} & FPS$\uparrow$ & \#Inst.($10^6$)$\downarrow$ & Mem.(MB)$\downarrow$ & PSNR$\uparrow$ & SSIM$\uparrow$ & \multicolumn{1}{c|}{LPIPS$\downarrow$} & FPS$\uparrow$ & \#Inst.($10^6$)$\downarrow$ & Mem.(MB)$\downarrow$ \\ 
\midrule[0.05em]
3DGS~\cite{kerbl20233d} & 27.46 & 0.812 & 0.223 & 20.79 & 2168.68 & 710.6 & 23.75 & 0.848 & 0.176 & 41.97 & 1034.37 & 430.9 & 29.59 & 0.902 & 0.244 & 23.94 & 1645.11 & 639.9 \\
3DGS+Opti. & 27.46 & 0.812 & 0.223 & 21.75 & 1945.88 & 710.6 & 23.75 & 0.847 & 0.176 & 45.96 & 923.51 & 430.9 & 29.59 & 0.902 & 0.244 & 29.60 & 1485.44 & 639.9 \\
3DGS+Opti.+$\cHP$ & 27.70 & 0.814 & 0.219 & 46.15 & 814.74 & 380.9 & 24.05 & 0.852 & 0.171 & 118.61 & 374.63 & 207.6 & 29.74 & 0.903 & 0.242 & 81.73 & \green{412.50} & 400.6 \\
3DGS+Opti.+$\cCR$ & 27.50 & 0.812 & 0.221 & 30.10 & 1574.92 & 710.6 & 23.70 & 0.848 & 0.173 & 57.03 & 763.83 & 430.9 & 29.70 & 0.902 & 0.242 & 38.79 & 1156.46 & 639.9 \\
\proj+ 3DGS & \green{27.72} & \green{0.814} & \green{0.217} & \green{59.67} & \green{778.25} & \green{380.9} & \green{24.02} & \green{0.852} & \green{0.168} & \green{127.80} & \green{356.15} & \green{207.6} & \green{29.79} & \green{0.903} & \green{0.240} & \green{86.58} & 421.74 & \green{400.6} \\
\bottomrule[0.15em]
\end{tabular}
}
\label{tab:abl}
\end{table*}

In this section, we show the contributions of individual optimizations.
Specifically, we evaluate four variants:
\begin{itemize}
    \item \mode{+Opti.}: this variant only includes the additional code optimizations proposed in \Sect{sec:eval:exp}.
    \item \mode{+Opti.+$\cHP$}: this variant includes both the code optimizations in \Sect{sec:eval:exp} and \textit{hybrid preprocessing} in \Sect{sec:method:hp}.
    \item \mode{+Opti.+$\cCR$}: this variant includes code optimizations in \Sect{sec:eval:exp} and \textit{contribution-aware rasterization} in \Sect{sec:method:cr}.
    \item \mode{\proj}: this variant is the full-fledged algorithm.
\end{itemize}

\Tbl{tab:abl} presents the results of our ablation study on \mode{3DGS}.
Across all datasets, our code optimization (\mode{+Opti.}) achieves a 1.1$\times$ speedup.
Building on this, \mode{+Opti.+$\cHP$} and \mode{+Opti.+$\cCR$} yield additional 2.8$\times$ and 1.3$\times$ speedups, respectively.
When all optimizations are combined, \mode{\proj} achieves an overall 3.2$\times$ speedup.
All model weight savings come from $\cHP$, as the other two optimizations do not contribute to the model reduction.
Note that, \mode{\proj} may increase the overall instruction count but improves the overall speedup due to the reductions on warp divergence.

In terms of the rendering quality, the additional optimizations proposed in \Sect{sec:eval:exp} do not change the accuracy.
The $\cHP$ technique from \Sect{sec:method:hp}, combined with fine-tuning as described in \Sect{sec:method:ct}, further enhances rendering quality across all three quality metrics.
On average, $\cHP$ improves the rendering quality by 0.23 on PSNR.
The effect of $\cCR$ from \Sect{sec:method:cr} on quality is minimal, 0.03 in PSNR.

\begin{table} 
\centering
\caption{Quality and performance comparison on \mode{3DGS} with and without integrated fine-tuning on \mode{3DGS}.}
\resizebox{\linewidth}{!}{
\renewcommand*{\arraystretch}{1}
\renewcommand*{\tabcolsep}{5pt}
\begin{tabular}{ c|cccccc } 
\toprule[0.15em]
Dataset & \multicolumn{5}{c}{\textbf{Mip-NeRF360}} \\ 
Metric & PSNR$\uparrow$ & SSIM$\uparrow$ & LPIPS$\downarrow$ & FPS$\uparrow$ & \#Inst.($10^6$)$\downarrow$ & Mem. (MB) $\downarrow$\\ 
\midrule[0.05em]
w/o Fine-tuning & 27.48 & 0.812 & 0.221 & 59.72 & 768.76 & 380.9 \\
w/ Fine-tuning & 27.72 & 0.814 & 0.217 & 59.67 & 778.25 & 380.9 \\
\bottomrule[0.15em]
\end{tabular}
}
\label{tab:cotrain}
\end{table}

\begin{table} 
\caption{Ablation study on $\cL_{\text{consistency}}$. Lower is better.}
\centering
\resizebox{\linewidth}{!}{
\renewcommand*{\arraystretch}{1}
\renewcommand*{\tabcolsep}{3pt}
\begin{tabular}{ c|cc|cc|cc } 
\toprule[0.15em]
Dataset & \multicolumn{2}{c|}{\textbf{Mip-NeRF360}} & \multicolumn{2}{c|}{\textbf{Tanks\&Temples}} & \multicolumn{2}{c}{\textbf{Deep Blending}} \\ 
& w/ Loss & w/o Loss & w/ Loss & w/o Loss & w/ Loss & w/o Loss \\ 
\midrule[0.05em]
3DGS & \green{0.058} & 0.060 & \green{0.067} & 0.074 & \green{0.042} & 0.043 \\
\bottomrule[0.15em]
\end{tabular}
}
\label{tab:loss_flip}
\end{table}

\paragraph{Fine-Tuning.}
Next, we present the results with and without the integrated fine-tuning, as described in \Sect{sec:method:ct}.
Here, we focus on the results of the Mip-NeRF360 dataset with \mode{3DGS}.
Similar trends are observed across other datasets in the supplementary.
\Tbl{tab:cotrain} shows that fine-tuning results in significant improvements.
Meanwhile, fine-tuning does not affect performance for both FPS and GPU memory consumption.
\Tbl{tab:loss_flip} also shows the effectiveness of $\cL_{\text{consistency}}$, finetuning with our proposed loss term can improve the rendering quality.

\begin{figure}
  \centering
  \begin{subfigure}{0.48\linewidth}
    \includegraphics[width=\columnwidth]{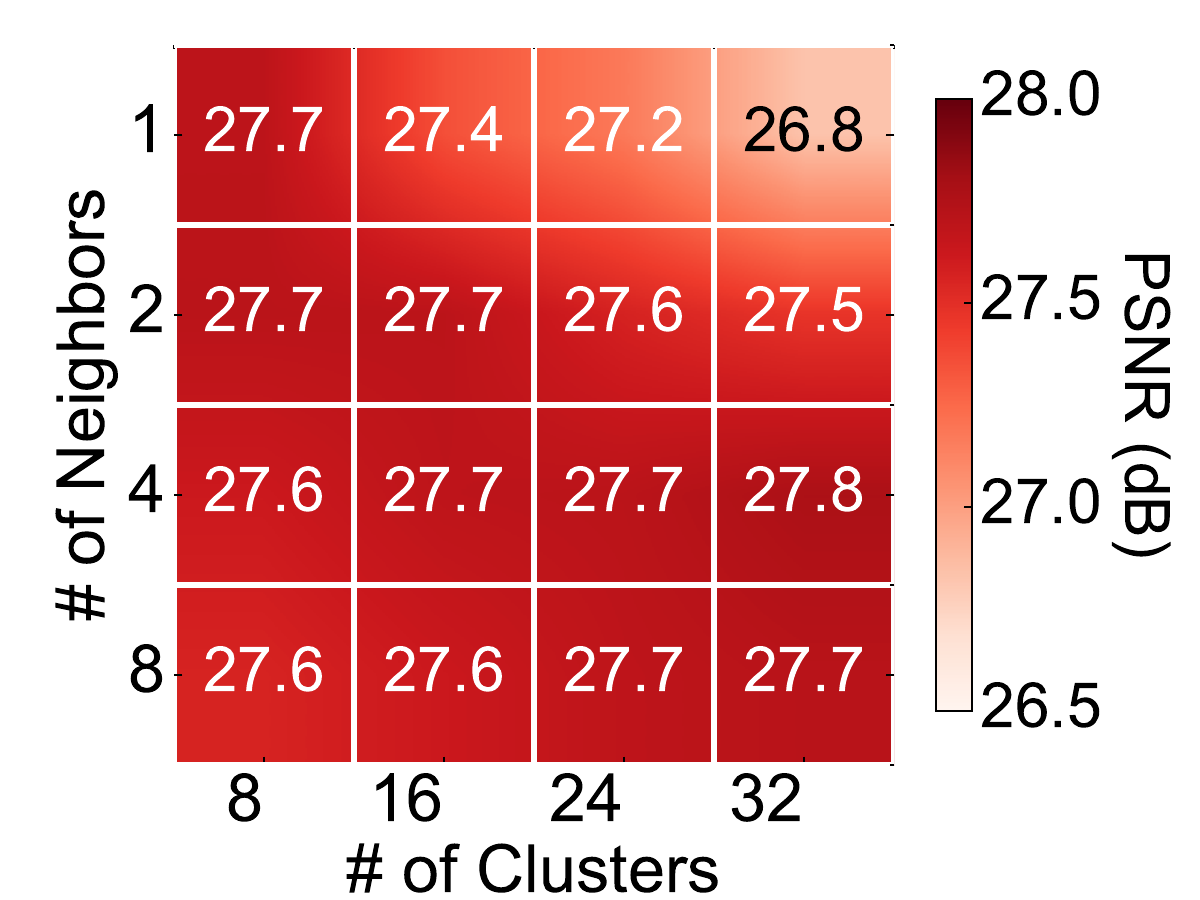}
    \caption{PSNR.}
    \label{fig:cluster_acc}
  \end{subfigure}
  \hfill
  \begin{subfigure}{0.48\linewidth}
    \includegraphics[width=\columnwidth]{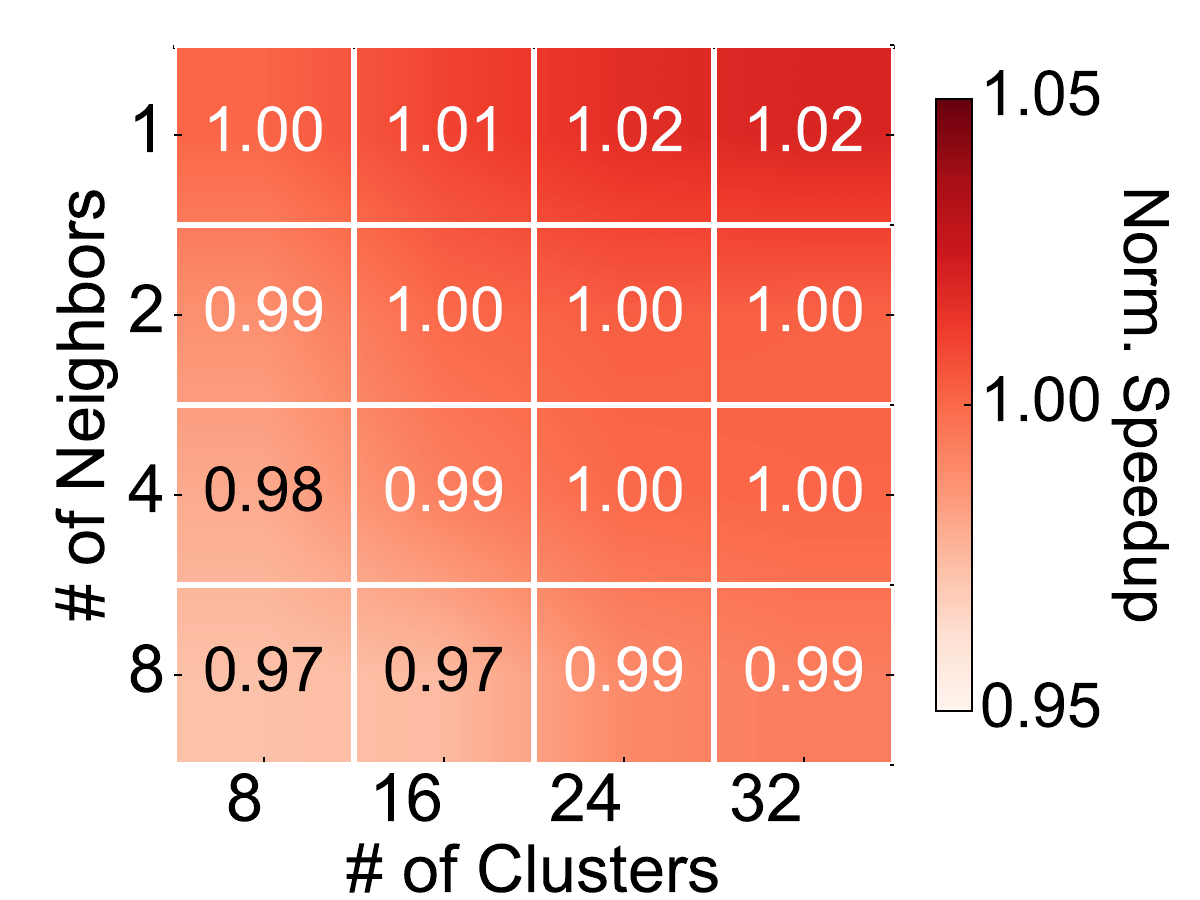}
    \caption{Normalized speedup.}
    \label{fig:cluster_perf}
  \end{subfigure}
  \caption{Sensitivity of rendering quality and performance to the number of clusters and cluster neighbors in \Sect{sec:method:hp}.}
  \label{fig:cluster}
\end{figure}

\subsection{Sensitivity Study}
\label{sec:eval:sens}

\paragraph{Number of Clusters.}
\Fig{fig:cluster} shows how rendering quality and performance vary with two key hyperparameters, the number of clusters and the number of neighboring clusters during rendering ($M$), in \Sect{sec:method:hp}.
Performance numbers are normalized to our default setting of 24 clusters with 4 neighbors (including itself). 
As the number of neighboring clusters increases, speedup decreases, while accuracy initially improves. 
However, too many clusters, e.g., 8 clusters with 8 neighbors, degrades accuracy. 
This degradation is because each Gaussian is responsible for too many view directions, negatively impacting rendering quality.

\begin{figure}[t]
\centering
\begin{minipage}[t]{0.48\columnwidth}
  \centering
  \includegraphics[width=\columnwidth]{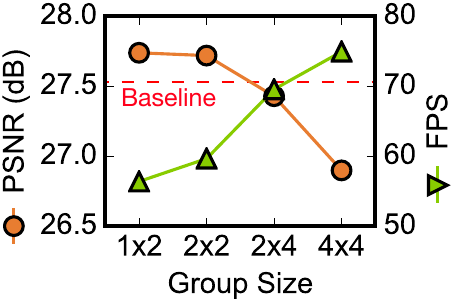}
  \caption{Sensitivity of rendering quality and performance to the pixel group size.}
  \label{fig:window_size_sens}
\end{minipage}
\hspace{2pt}
\begin{minipage}[t]{0.48\columnwidth}
  \centering
  \includegraphics[width=\columnwidth]{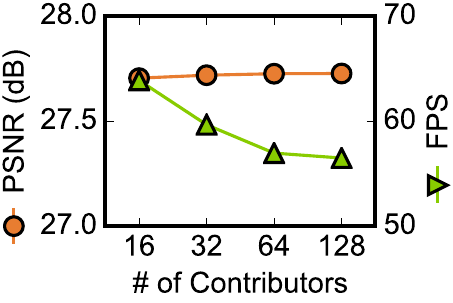}
  \caption{Sensitivity of rendering quality and performance to the number of contributors.}
  \label{fig:contributor_sens}
\end{minipage}
\end{figure}

\paragraph{Pixel Group.}
\Fig{fig:window_size_sens} illustrates the sensitivity of both the rendering quality and performance to the pixel group size using Mip-NeRF360~\cite{barron2021mip}, as defined in \Sect{sec:method:cr}.
Other datasets show a similar trend.
As the pixel group size increases, rendering quality initially decreases slightly, but the drop becomes more significant once the group size exceeds $2\times 4$.
% Nevertheless, we can achieve better rendering quality when the group size is smaller than $2 \times 4$.
% We also present examples of rendered images with different group sizes in \Fig{fig:ws_example}.
% As the group size increases to $2\times 4$, artifacts begin to appear in the high-frequency regions.
% On the performance side, speedup increases sublinearly at first, but it almost plateaus once the group size reaches $2\times 4$.
% Thus, we ultimately select $2\times 2$.
By trading off between quality and performance, we ultimately select the pixel group size of $2\times 2$.

\paragraph{Number of Contributors.} \Fig{fig:contributor_sens} shows the sensitivity of the rendering quality and performance to the number of primary contributors in the clustering strategy (\Sect{sec:method:hp}).
The rendering quality is not significantly affected by the number of contributors, while the performance gradually decreases as the number of contributors increases.
\section{Conclusion}
\label{sec:disc}

Recent neural rendering has transformed the landscape of real-time rendering.
3DGS emerges as a new rendering primitive that offers unprecedented realism using the conventional rasterization pipeline.
This paper proposes two principled and GPU-oriented optimizations that drastically improve the general Gaussian splatting pipelines by understanding and leveraging the GPUs' characteristics.
We believe that this work could be the first step to understanding the fundamental limitations of GPUs in supporting 3DGS.
The result would be valuable to rethink the next architectures tailored to this future technique.

{
    \small
    \bibliographystyle{ieeenat_fullname}
    \bibliography{main}

@String(TOG= {ACM Trans. Graph.})

@String(TOG   = {ACM TOG})

@article{ren2024octree,
  title={Octree-gs: Towards consistent real-time rendering with lod-structured 3d gaussians},
  author={Ren, Kerui and Jiang, Lihan and Lu, Tao and Yu, Mulin and Xu, Linning and Ni, Zhangkai and Dai, Bo},
  journal={arXiv preprint arXiv:2403.17898},
  year={2024}
}

@article{kerbl2024hierarchical,
  title={A hierarchical 3d gaussian representation for real-time rendering of very large datasets},
  author={Kerbl, Bernhard and Meuleman, Andreas and Kopanas, Georgios and Wimmer, Michael and Lanvin, Alexandre and Drettakis, George},
  journal={ACM Transactions on Graphics (TOG)},
  volume={43},
  number={4},
  pages={1--15},
  year={2024},
  publisher={ACM New York, NY, USA}
}

@article{fan2023lightgaussian,
  title={Lightgaussian: Unbounded 3d gaussian compression with 15x reduction and 200+ fps},
  author={Fan, Zhiwen and Wang, Kevin and Wen, Kairun and Zhu, Zehao and Xu, Dejia and Wang, Zhangyang},
  journal={arXiv preprint arXiv:2311.17245},
  year={2023}
}

@article{fang2024mini,
  title={Mini-Splatting: Representing Scenes with a Constrained Number of Gaussians},
  author={Fang, Guangchi and Wang, Bing},
  journal={arXiv preprint arXiv:2403.14166},
  year={2024}
}

@inproceedings{lee2024compact,
  title={Compact 3d gaussian representation for radiance field},
  author={Lee, Joo Chan and Rho, Daniel and Sun, Xiangyu and Ko, Jong Hwan and Park, Eunbyung},
  booktitle={Proceedings of the IEEE/CVF Conference on Computer Vision and Pattern Recognition},
  pages={21719--21728},
  year={2024}
}

@inproceedings{girish2025eagles,
  title={Eagles: Efficient accelerated 3d gaussians with lightweight encodings},
  author={Girish, Sharath and Gupta, Kamal and Shrivastava, Abhinav},
  booktitle={European Conference on Computer Vision},
  pages={54--71},
  year={2025},
  organization={Springer}
}

@article{niemeyer2024radsplat,
  title={Radsplat: Radiance field-informed gaussian splatting for robust real-time rendering with 900+ fps},
  author={Niemeyer, Michael and Manhardt, Fabian and Rakotosaona, Marie-Julie and Oechsle, Michael and Duckworth, Daniel and Gosula, Rama and Tateno, Keisuke and Bates, John and Kaeser, Dominik and Tombari, Federico},
  journal={arXiv preprint arXiv:2403.13806},
  year={2024}
}

@inproceedings{mallick2024taming,
  title={Taming 3dgs: High-quality radiance fields with limited resources},
  author={Mallick, Saswat Subhajyoti and Goel, Rahul and Kerbl, Bernhard and Steinberger, Markus and Carrasco, Francisco Vicente and De La Torre, Fernando},
  booktitle={SIGGRAPH Asia 2024 Conference Papers},
  pages={1--11},
  year={2024}
}

@inproceedings{wang2024adr,
  title={AdR-Gaussian: Accelerating Gaussian Splatting with Adaptive Radius},
  author={Wang, Xinzhe and Yi, Ran and Ma, Lizhuang},
  booktitle={SIGGRAPH Asia 2024 Conference Papers},
  pages={1--10},
  year={2024}
}

@article{gui2024balanced,
  title={Balanced 3DGS: Gaussian-wise Parallelism Rendering with Fine-Grained Tiling},
  author={Gui, Hao and Hu, Lin and Chen, Rui and Huang, Mingxiao and Yin, Yuxin and Yang, Jin and Wu, Yong},
  journal={arXiv preprint arXiv:2412.17378},
  year={2024}
}

@inproceedings{feng2025flashgs,
  title={Flashgs: Efficient 3d gaussian splatting for large-scale and high-resolution rendering},
  author={Feng, Guofeng and Chen, Siyan and Fu, Rong and Liao, Zimu and Wang, Yi and Liu, Tao and Hu, Boni and Xu, Linning and Pei, Zhilin and Li, Hengjie and others},
  booktitle={Proceedings of the Computer Vision and Pattern Recognition Conference},
  pages={26652--26662},
  year={2025}
}

@inproceedings{lee2024gscore,
  title={GSCore: Efficient Radiance Field Rendering via Architectural Support for 3D Gaussian Splatting},
  author={Lee, Junseo and Lee, Seokwon and Lee, Jungi and Park, Junyong and Sim, Jaewoong},
  booktitle={Proceedings of the 29th ACM International Conference on Architectural Support for Programming Languages and Operating Systems, Volume 3},
  pages={497--511},
  year={2024}
}

@article{lin2024rtgs,
  title={RTGS: Enabling Real-Time Gaussian Splatting on Mobile Devices Using Efficiency-Guided Pruning and Foveated Rendering},
  author={Lin, Weikai and Feng, Yu and Zhu, Yuhao},
  journal={arXiv preprint arXiv:2407.00435},
  year={2024}
}

@inproceedings{pei2025gcc,
  title={GCC: A 3DGS Inference Architecture with Gaussian-Wise and Cross-Stage Conditional Processing},
  author={Pei, Minnan and Li, Gang and Si, Junwen and Zhu, Zeyu and Mo, Zitao and Wang, Peisong and Song, Zhuoran and Liang, Xiaoyao and Cheng, Jian},
  booktitle={Proceedings of the 58th IEEE/ACM International Symposium on Microarchitecture{\textregistered}},
  pages={1824--1837},
  year={2025}
}

@article{kerbl20233d,
  title={3d gaussian splatting for real-time radiance field rendering},
  author={Kerbl, Bernhard and Kopanas, Georgios and Leimk{\"u}hler, Thomas and Drettakis, George},
  journal={ACM Transactions on Graphics},
  volume={42},
  number={4},
  pages={1--14},
  year={2023},
  publisher={ACM}
}

@inproceedings{barron2021mip,
  title={Mip-nerf: A multiscale representation for anti-aliasing neural radiance fields},
  author={Barron, Jonathan T and Mildenhall, Ben and Tancik, Matthew and Hedman, Peter and Martin-Brualla, Ricardo and Srinivasan, Pratul P},
  booktitle={Proceedings of the IEEE/CVF International Conference on Computer Vision},
  pages={5855--5864},
  year={2021}
}

@article{liu2024citygaussianv2,
  title={CityGaussianV2: Efficient and Geometrically Accurate Reconstruction for Large-Scale Scenes},
  author={Liu, Yang and Luo, Chuanchen and Mao, Zhongkai and Peng, Junran and Zhang, Zhaoxiang},
  journal={arXiv preprint arXiv:2411.00771},
  year={2024}
}

@inproceedings{liu2025citygaussian,
  title={Citygaussian: Real-time high-quality large-scale scene rendering with gaussians},
  author={Liu, Yang and Luo, Chuanchen and Fan, Lue and Wang, Naiyan and Peng, Junran and Zhang, Zhaoxiang},
  booktitle={European Conference on Computer Vision},
  pages={265--282},
  year={2025},
  organization={Springer}
}

@inproceedings{tancik2022block,
  title={Block-nerf: Scalable large scene neural view synthesis},
  author={Tancik, Matthew and Casser, Vincent and Yan, Xinchen and Pradhan, Sabeek and Mildenhall, Ben and Srinivasan, Pratul P and Barron, Jonathan T and Kretzschmar, Henrik},
  booktitle={Proceedings of the IEEE/CVF Conference on Computer Vision and Pattern Recognition},
  pages={8248--8258},
  year={2022}
}

@inproceedings{rojas2023re,
  title={Re-ReND: Real-time Rendering of NeRFs across Devices},
  author={Rojas, Sara and Zarzar, Jesus and P{\'e}rez, Juan C and Sanakoyeu, Artsiom and Thabet, Ali and Pumarola, Albert and Ghanem, Bernard},
  booktitle={Proceedings of the IEEE/CVF International Conference on Computer Vision},
  pages={3632--3641},
  year={2023}
}

@inproceedings{chen2023mobilenerf,
  title={Mobilenerf: Exploiting the polygon rasterization pipeline for efficient neural field rendering on mobile architectures},
  author={Chen, Zhiqin and Funkhouser, Thomas and Hedman, Peter and Tagliasacchi, Andrea},
  booktitle={Proceedings of the IEEE/CVF Conference on Computer Vision and Pattern Recognition},
  pages={16569--16578},
  year={2023}
}

@inproceedings{hedman2021baking,
  title={Baking neural radiance fields for real-time view synthesis},
  author={Hedman, Peter and Srinivasan, Pratul P and Mildenhall, Ben and Barron, Jonathan T and Debevec, Paul},
  booktitle={Proceedings of the IEEE/CVF International Conference on Computer Vision},
  pages={5875--5884},
  year={2021}
}

@inproceedings{hu2022efficientnerf,
  title={Efficientnerf efficient neural radiance fields},
  author={Hu, Tao and Liu, Shu and Chen, Yilun and Shen, Tiancheng and Jia, Jiaya},
  booktitle={Proceedings of the IEEE/CVF Conference on Computer Vision and Pattern Recognition},
  pages={12902--12911},
  year={2022}
}

@inproceedings{zwicker2001ewa,
  title={EWA volume splatting},
  author={Zwicker, Matthias and Pfister, Hanspeter and Van Baar, Jeroen and Gross, Markus},
  booktitle={Proceedings Visualization, 2001. VIS'01.},
  pages={29--538},
  year={2001},
  organization={IEEE}
}

@article{feng2024cicero,
  title={Cicero: Addressing Algorithmic and Architectural Bottlenecks in Neural Rendering by Radiance Warping and Memory Optimizations},
  author={Feng, Yu and Liu, Zihan and Leng, Jingwen and Guo, Minyi and Zhu, Yuhao},
  journal={arXiv preprint arXiv:2404.11852},
  year={2024}
}

@article{hou2020motion,
  title={Motion prediction and pre-rendering at the edge to enable ultra-low latency mobile 6DoF experiences},
  author={Hou, Xueshi and Dey, Sujit},
  journal={IEEE Open Journal of the Communications Society},
  volume={1},
  pages={1674--1690},
  year={2020},
  publisher={IEEE}
}

@article{hedman2018deep,
  title={Deep blending for free-viewpoint image-based rendering},
  author={Hedman, Peter and Philip, Julien and Price, True and Frahm, Jan-Michael and Drettakis, George and Brostow, Gabriel},
  journal={ACM Transactions on Graphics (ToG)},
  volume={37},
  number={6},
  pages={1--15},
  year={2018},
  publisher={ACM New York, NY, USA}
}

@article{knapitsch2017,
    author    = {Arno Knapitsch and Jaesik Park and Qian-Yi Zhou and Vladlen Koltun},
    title     = {Tanks and Temples: Benchmarking Large-Scale Scene Reconstruction},
    journal   = {ACM Transactions on Graphics},
    volume    = {36},
    number    = {4},
    year      = {2017},
}

@inproceedings{gottschalk1996obbtree,
  title={OBBTree: A hierarchical structure for rapid interference detection},
  author={Gottschalk, Stefan and Lin, Ming C and Manocha, Dinesh},
  booktitle={Proceedings of the 23rd annual conference on Computer graphics and interactive techniques},
  pages={171--180},
  year={1996}
}

@article{klosowski1998efficient,
  title={Efficient collision detection using bounding volume hierarchies of k-DOPs},
  author={Klosowski, James T and Held, Martin and Mitchell, Joseph SB and Sowizral, Henry and Zikan, Karel},
  journal={IEEE Transactions on Visualization and Computer Graphics},
  volume={4},
  number={1},
  pages={21--36},
  year={1998},
  publisher={IEEE}
}

@misc{a6000,
  title={NVIDIA RTX A6000},
  url={https://www.techpowerup.com/gpu-specs/rtx-a6000.c3686}
}

@misc{orin_nx,
  title={NVIDIA Jetson Orin NX 16 GB},
  url={https://www.techpowerup.com/gpu-specs/jetson-orin-nx-16-gb.c4086}
}

@article{wang2023effect,
  title={Effect of frame rate on user experience, performance, and simulator sickness in virtual reality},
  author={Wang, Jialin and Shi, Rongkai and Zheng, Wenxuan and Xie, Weijie and Kao, Dominic and Liang, Hai-Ning},
  journal={IEEE Transactions on Visualization and Computer Graphics},
  volume={29},
  number={5},
  pages={2478--2488},
  year={2023},
  publisher={IEEE}
}

@inproceedings{lin2025metasapiens,
  title={MetaSapiens: Real-Time Neural Rendering with Efficiency-Aware Pruning and Accelerated Foveated Rendering},
  author={Lin, Weikai and Feng, Yu and Zhu, Yuhao},
  booktitle={Proceedings of the 30th ACM International Conference on Architectural Support for Programming Languages and Operating Systems, Volume 1},
  pages={669--682},
  year={2025}
}

@article{andersson2020flip,
  title={FLIP: A Difference Evaluator for Alternating Images.},
  author={Andersson, Pontus and Nilsson, Jim and Akenine-M{\"o}ller, Tomas and Oskarsson, Magnus and {\AA}str{\"o}m, Kalle and Fairchild, Mark D},
  journal={Proc. ACM Comput. Graph. Interact. Tech.},
  volume={3},
  number={2},
  pages={15--1},
  year={2020}
}

@inproceedings{hanson2025speedy,
  title={Speedy-splat: Fast 3d gaussian splatting with sparse pixels and sparse primitives},
  author={Hanson, Alex and Tu, Allen and Lin, Geng and Singla, Vasu and Zwicker, Matthias and Goldstein, Tom},
  booktitle={Proceedings of the Computer Vision and Pattern Recognition Conference},
  pages={21537--21546},
  year={2025}
}

@inproceedings{hanson2025pup,
  title={Pup 3d-gs: Principled uncertainty pruning for 3d gaussian splatting},
  author={Hanson, Alex and Tu, Allen and Singla, Vasu and Jayawardhana, Mayuka and Zwicker, Matthias and Goldstein, Tom},
  booktitle={Proceedings of the Computer Vision and Pattern Recognition Conference},
  pages={5949--5958},
  year={2025}
}

@inproceedings{liu2025maskgaussian,
  title={Maskgaussian: Adaptive 3d gaussian representation from probabilistic masks},
  author={Liu, Yifei and Zhong, Zhihang and Zhan, Yifan and Xu, Sheng and Sun, Xiao},
  booktitle={Proceedings of the Computer Vision and Pattern Recognition Conference},
  pages={681--690},
  year={2025}
}

@article{taghipour2025svr,
  title={SVR-GS: Spatially Variant Regularization for Probabilistic Masks in 3D Gaussian Splatting},
  author={Taghipour, Ashkan and Naghshin, Vahid and Southwell, Benjamin and Boussaid, Farid and Laga, Hamid and Bennamoun, Mohammed},
  journal={arXiv preprint arXiv:2509.11116},
  year={2025}
}

@inproceedings{niedermayr2024compressed,
  title={Compressed 3d gaussian splatting for accelerated novel view synthesis},
  author={Niedermayr, Simon and Stumpfegger, Josef and Westermann, R{\"u}diger},
  booktitle={Proceedings of the IEEE/CVF Conference on Computer Vision and Pattern Recognition},
  pages={10349--10358},
  year={2024}
}

@inproceedings{chen2024hac,
  title={Hac: Hash-grid assisted context for 3d gaussian splatting compression},
  author={Chen, Yihang and Wu, Qianyi and Lin, Weiyao and Harandi, Mehrtash and Cai, Jianfei},
  booktitle={European Conference on Computer Vision},
  pages={422--438},
  year={2024},
  organization={Springer}
}

@article{chen2025hac++,
  title={Hac++: Towards 100x compression of 3d gaussian splatting},
  author={Chen, Yihang and Wu, Qianyi and Lin, Weiyao and Harandi, Mehrtash and Cai, Jianfei},
  journal={arXiv preprint arXiv:2501.12255},
  year={2025}
}

@article{papantonakis2024reducing,
  title={Reducing the memory footprint of 3d gaussian splatting},
  author={Papantonakis, Panagiotis and Kopanas, Georgios and Kerbl, Bernhard and Lanvin, Alexandre and Drettakis, George},
  journal={Proceedings of the ACM on Computer Graphics and Interactive Techniques},
  volume={7},
  number={1},
  pages={1--17},
  year={2024},
  publisher={ACM New York, NY, USA}
}

@article{zhang2024lp,
  title={Lp-3dgs: Learning to prune 3d gaussian splatting},
  author={Zhang, Zhaoliang and Song, Tianchen and Lee, Yongjae and Yang, Li and Peng, Cheng and Chellappa, Rama and Fan, Deliang},
  journal={Advances in Neural Information Processing Systems},
  volume={37},
  pages={122434--122457},
  year={2024}
}

@inproceedings{durvasula2025contrags,
  title={ContraGS: Codebook-Condensed and Trainable Gaussian Splatting for Fast, Memory-Efficient Reconstruction},
  author={Durvasula, Sankeerth and Muhunthan, Sharanshangar and Moustafa, Zain and Chen, Richard and Liang, Ruofan and Guan, Yushi and Ahuja, Nilesh and Jain, Nilesh and Panneer, Selvakumar and Vijaykumar, Nandita},
  booktitle={Proceedings of the IEEE/CVF International Conference on Computer Vision},
  pages={28935--28945},
  year={2025}
}

@article{fang2024mini2,
  title={Mini-splatting2: Building 360 scenes within minutes via aggressive gaussian densification},
  author={Fang, Guangchi and Wang, Bing},
  journal={arXiv preprint arXiv:2411.12788},
  year={2024}
}

@inproceedings{hamdi2024ges,
  title={Ges: Generalized exponential splatting for efficient radiance field rendering},
  author={Hamdi, Abdullah and Melas-Kyriazi, Luke and Mai, Jinjie and Qian, Guocheng and Liu, Ruoshi and Vondrick, Carl and Ghanem, Bernard and Vedaldi, Andrea},
  booktitle={Proceedings of the IEEE/CVF Conference on Computer Vision and Pattern Recognition},
  pages={19812--19822},
  year={2024}
}

@inproceedings{held20253d,
  title={3D convex splatting: Radiance field rendering with 3D smooth convexes},
  author={Held, Jan and Vandeghen, Renaud and Hamdi, Abdullah and Deliege, Adrien and Cioppa, Anthony and Giancola, Silvio and Vedaldi, Andrea and Ghanem, Bernard and Van Droogenbroeck, Marc},
  booktitle={Proceedings of the Computer Vision and Pattern Recognition Conference},
  pages={21360--21369},
  year={2025}
}

@inproceedings{li20253d,
  title={3D-HGS: 3D Half-Gaussian Splatting},
  author={Li, Haolin and Liu, Jinyang and Sznaier, Mario and Camps, Octavia},
  booktitle={Proceedings of the Computer Vision and Pattern Recognition Conference},
  pages={10996--11005},
  year={2025}
}

@article{ye2025gaussian,
  title={When gaussian meets surfel: Ultra-fast high-fidelity radiance field rendering},
  author={Ye, Keyang and Shao, Tianjia and Zhou, Kun},
  journal={ACM Transactions on Graphics (TOG)},
  volume={44},
  number={4},
  pages={1--15},
  year={2025},
  publisher={ACM New York, NY, USA}
}

@inproceedings{lee2025vr,
  title={VR-Pipe: Streamlining Hardware Graphics Pipeline for Volume Rendering},
  author={Lee, Junseo and Kim, Jaisung and Park, Junyong and Sim, Jaewoong},
  booktitle={2025 IEEE International Symposium on High Performance Computer Architecture (HPCA)},
  pages={217--230},
  year={2025},
  organization={IEEE}
}

@article{liao2025tc,
  title={Tc-gs: A faster gaussian splatting module utilizing tensor cores},
  author={Liao, Zimu and Ding, Jifeng and Cui, Siwei and Gong, Ruixuan and Hu, Boni and Wang, Yi and Li, Hengjie and Zhang, XIngcheng and Wang, Hui and Fu, Rong},
  journal={arXiv preprint arXiv:2505.24796},
  year={2025}
}
}

% WARNING: do not forget to delete the supplementary pages from your submission 
% \input{sec/X_suppl}

\end{document}